\documentclass[reqno,xcolor=dvipsnames]{amsart}
\usepackage[dvipsnames]{xcolor}

\usepackage{geometry}
\usepackage[round,comma]{natbib}                %

\usepackage{enumerate}
\usepackage{mathabx}

\usepackage{amsmath}
\usepackage{amsthm}
\usepackage{amsfonts}
\usepackage{amssymb}
\usepackage{dsfont}

\usepackage{nicefrac}
\usepackage{booktabs}
\usepackage{mathrsfs}
\usepackage{bm}
\usepackage{mathtools}

\newcommand{\ccB}{{\mathscr B}}

\newcommand{\ccF}{{\mathscr F}}

\newcommand{\ccG}{{\mathscr G}}

\newcommand{\ccH}{{\mathscr H}}

\newcommand{\kI}{{\mathfrak I}}

\newcommand{\kJ}{{\mathfrak J}}
\newcommand{\ccK}{{\mathscr K}}

\newcommand{\ccM}{{\mathscr M}}
\newcommand{\kM}{{\mathfrak M}}

\newcommand{\kN}{{\mathfrak N}}

\newcommand{\ccP}{{\mathscr P}}

\DeclareMathOperator{\Var}{Var}

\newcommand{\Ind}{{\mathds 1}}
\newcommand{\ind}[1]{\Ind_{\{#1\}}}

\newcommand{\restr}{\mathbf{\kern0.3ex%
 \vert\kern-0.3ex}\backprime\kern0.3ex}
\newcommand{\abs}[1]{|#1|}
\newcommand{\half}{\ ^1\mkern -3mu/ _2\ }
\newcommand{\threehalf}{\ ^3\mkern -3mu/ _2\ }
\newcommand{\norm}[1]{\|#1\|}

\newcommand{\FF}{\mathbb{F}}
\newcommand{\GG}{\mathbb{G}}
\newcommand{\HH}{\mathbb{H}}

\newcommand{\NN}{\mathbb{N}}

\newcommand{\RR}{\mathbb{R}}
\newcommand{\TT}{\mathbb{T}}

\newtheorem{theorem}{Theorem}[section]
\newtheorem{corollary}[theorem]{Corollary}      %
\newtheorem{proposition}[theorem]{Proposition}  %

\theoremstyle{definition}
\newtheorem{example}[theorem]{Example} %
\newtheorem{definition}[theorem]{Definition} %
\newtheorem{remark}[theorem]{Remark}%
\newtheorem{assumption}[theorem]{Assumption}%

\usepackage[backgroundcolor=white,bordercolor=orange]{todonotes}

\usepackage[colorlinks,urlcolor=red,citecolor=blue,linkcolor=red]{hyperref}

\newcommand{\QcirP}{\operatorname{\Q\mkern-2mu\odot\mkern-2mu\P}}
\newcommand{\subQcirP}{\operatorname{\Q \odot \P}}
\newcommand{\subQetacirP}{\operatorname{\Q^\eta \odot \P}}

\newcommand{\nohbr}[1]{^{#1}}

\DeclareMathOperator{\esssup}{ess  \sup}
\DeclareMathOperator{\essinf}{ess  \inf}

\usepackage{subfiles} %

\makeatletter
\def\namedlabel#1#2{\begingroup
    #2%
    \def\@currentlabel{#2}%
    \phantomsection\label{#1}\endgroup
}
\makeatother

\usepackage{savesym}
\savesymbol{origcomment} %
\usepackage{changes} %
\definechangesauthor[name = Thorsten Schmidt, color=PineGreen]{TS}  %
\definechangesauthor[name = Karl-Theodor, color=magenta]{KT}

\newcommand{\bbG}{\GG}

\DeclareMathOperator{\DB}{DB}
\DeclareMathOperator{\SB}{SB}
\DeclareMathOperator{\AB}{AB}

\begin{document}

\renewcommand{\P}{P}
\newcommand{\Q}{Q}
\renewcommand{\theenumi}{\roman{enumi}}

\title{Insurance-Finance Arbitrage}
		\author[Artzner]{Philippe Artzner}
		\author[Eisele]{Karl-Theodor Eisele}
		\author[Schmidt]{Thorsten Schmidt}
		\address{Albert-Ludwigs University of Freiburg, Ernst-Zermelo-Str. 1, 79104 Freiburg, Germany.}
		\email{ thorsten.schmidt@stochastik.uni-freiburg.de}
		\address{University of Strasbourg, 7 rue Ren\'{e}  Descartes, Strasbourg 67084, France.}
    \email { artzner@math.unistra.fr}
    \email{ eisele@unistra.fr}
    \date{\today. }
    \thanks{The authors gratefully acknowledge the financial support from the Freiburg Institute for Advanced Studies (FRIAS) and the University of Strasbourg Institute of Advances Studies (USIAS) within the FRIAS-USIAS Research Project `Linking Finance and Insurance: Theory and Applications (2017--2019). Support from the DFG in the project SCHM 2160/15-1 and from the Freiburg Center for Data Analysis and Modeling (FDM) is gratefully acknowledged. Moreover, we would like to thank David Criens, Katharina Oberpriller, Lars Niemann and Moritz Ritter for stimulating discussions and their help with this project.}

\begin{abstract}
Most insurance contracts are inherently linked to financial markets, be it via interest rates, or -- as hybrid products like equity-linked life insurance and variable annuities -- directly to stocks or indices. However, insurance contracts are not for trade except sometimes as surrender to the selling office. This excludes the situation of arbitrage by buying and selling insurance contracts at different prices.   
Furthermore, the insurer uses private information on top of the publicly available one about financial market.
This paper provides a study of the consistency of insurance contracts in connection with trades in the financial market with explicit mention of the information involved.

By defining strategies on an insurance portfolio and combining them with financial trading strategies, we arrive at the notion of insurance-finance arbitrage (IFA). In analogy to the classical fundamental theorem of asset pricing, we give a fundamental theorem on the absence of IFA, leading to the existence of an insurance-finance-consistent probability. In addition, we study when this probability  gives the expected discounted cash-flows required by the EIOPA best estimate. 

The generality of our approach allows to incorporate many important aspects, like mortality risk or general levels of dependence between mortality and stock markets.  Utilizing the theory of enlargements of filtrations, we construct a tractable framework for insurance-finance consistent valuation. 
\bigskip

\end{abstract}

\maketitle

\section{Introduction}

An insurance contract and a financial asset both qualify as \emph{securities}, defined as \emph{the legal representation of the right to receive prospective future benefits under stated conditions} (\cite{sharpe1999investments}). But an insurance contract differs substantially from a financial security since it can be sold only by a regulated office to an individual appearing explicitly in the contract (see for example \cite{Boudreault2019}). In general, the owner cannot resell the contract (with some exceptions of surrender possibilities in life insurance). He can not benefit from simultaneously high and low prices (premium) for the same contract as in arbitrage facilities in financial markets  (\cite{sharpe1999investments}). On the other hand, the insurer might have arbitrage occasions by combining selling of contracts and trading in the market. In this case, a novel version of the arbitrage property has to be defined, which should at the same time incorporate the \emph{pooling principle} as the fundamental tool in insurance. 

It is a central observation that nowadays insurance contracts are often inherently and systematically linked to financial markets, be it via interest rates, or via direct links of the contractual benefits to stocks or indices (see \cite{Dhaenekukushetal-2013}). The associated risk can be mitigated by trading on financial markets. Therefore, we have to blend finance and insurance notions. 
On the insurance side, no re-balancing takes place and insurance companies generally form a large portfolio of homogeneous contracts for pooling and reducing risks. These considerations have to be taken into account by a generalization of the existing notion of arbitrage in a financial market.

Another crucial difference between finance an insurance is the necessity of enlarging the filtration about macroeconomic data (including market prices as well as globally available information for insurance firms like mortality tables) with privately held information on the population of the individual policyholders.

The mentioned differences prevent a simple transfer of arbitrage to the insurance situation and we propose a new concept of an \emph{insurance-finance arbitrage (IFA)}. Formally, it is natural to express the arbitrage of insurance flows including those of financial trading, by looking at a  with-probability-one event resulting in a net gain. This is the first point of the paper.

The apotheosis of mathematical finance is its Fundamental Theorem of Asset Pricing (FTAP) where, assuming no financial arbitrage, discounted prices of traded securities, given out of the blue, are explained as expectations. In discrete time it is known as the Dalang-Morton-Willinger Theorem (see \cite{dalang-etal1990}, \cite{schachermayer1992}, \cite{CarassusPhamTouzi2001}, \cite{DelbaenSchachermayer2006}, \cite{KabanovStricker2006}, to mention only few). Under the necessary extension adapted to the insurance situation we provide -- as first main result -- the corresponding Theorem \ref{thm-FTIFA}. To the best of our knowledge, it is the first time that the Theorem couples the pooling method in insurance with the financial no-arbitrage principle. On the one side,  absence of IFA implies the existence of an equivalent probability measure which, restricted to the financial market, is a martingale measure, and satisfies a super-martingale inequality on the insurance side. 
We call such a measure \emph{insurance-finance-consistent}.  The super-martingale property  resembles naturally the FTAP under short sales prohibition (see \cite{Pulido2014}).  On the other side, the fact that the possibility of an insurance-finance arbitrage involves two different kinds of strategies --- the market trading and the selling of insurance contracts --- has the consequence that the mere existence of an equivalent supermartingale measure is not sufficient to ensure absence of IFA; bounded insurance portfolio strategies  together with an additional condition (Assumption \ref{AssX1*}) are needed to do so.

In contrast to the classical situation on financial markets where finance-consistent (or risk neutral) probabilities are used to evaluate contingent claims, a general insurance-finance-consistent measure is of no great use for the insurance company since it lacks the connection to the insurance's statistical data. 
Assumption \ref{AssX1*} ensures that the statistical probability $P$ is reflected in a suitable way. 
As a consequence, we arrive at a class of measures which combine a martingale measure $Q$ on the market filtration with the statistical probability $P$ derived from insurance's internal information.
 This construction will be called the QP-rule. The result of it appears in \cite{plachky1984conservation}, studying measure extensions in a statistical context. The same problem arises in the context of time-consistent dynamic risk measures, see for example \cite{cheridito-etal-2006}.
Already \cite{dybvig1992hedging} proposed to use this tool for the evaluation of non-traded wealth  in the framework of state-price densities. It  also appears in \cite{PelsserStadje2014} in the case where linear rules are used for valuation.

In our opinion, the measure $\QcirP$ is the one to be used to compute the \emph{best estimate of liabilities (BEL)}, legally imposed by the European Insurance and Occupational Pensions Authority (EIOPA). The presence of a risk neutral measure $Q$ ensures moreover that the computed BEL is market consistent. However, with respect to EIOPA's supervision rules, we only looked at the first step of insurance solvency regulation, i.e.\ at the linear case. The following non-linear ones, like the risk margin (RM) and the solvency capital requirement (SCR), are left for future research.

The intertwining of the two measures $Q$ and $P$ in the QP-rule makes direct calculations difficult, in particular, when it comes to determine the insurance-finance consistency of it. A sufficient condition for IFA is given in our second result, Theorem \ref{thm-NIFA}. In a slightly different setting, Corollary \ref{col:NIFA} yields a sufficient and a necessary condition for the insurance-finance consistency of the QP-rule.  

Since our setting is very general, we are able to pass existing approaches: first, we consider a financial market without restricting to the complete case; second, we allow for arbitrary dependence between the financial market and the insurance quantities - this is important in many insurance products, for example when considering surrender behaviour or stochastic mortality.
Paradigmatically, the encountered pandemic highlights the necessity of  allowing  dependence between mortality and stock markets; third, we do not exclude financial arbitrage when trading were done with strategies adapted to the enlarged filtration but we restrict the insurance's hedging to strategies corresponding to the law against insider trading. Moreover, the insurance benefits need not to be bounded, as is often required when dealing with risk measures (see \cite{artzner1999coherent}, \cite{riedel2004dynamic},  \cite{cheridito-etal-2006}).  However, when classifying arbitrage possibilities, we distinguish between unbounded and bounded portfolio strategies (see Definition \ref{def-IFarb}).   

Finally, we exemplify the insurance-finance consistency of the QP-rule by a tractable result for a variable annuity contract whose benefits depend on the events of death and surrender. These random times produce events which give rise to a \emph{progressive enlargement} of the initial filtration and, nevertheless, can be evaluated by a nice formula.
The linkage between insurance contracts and financial markets is a problem which has been intensively studied in the literature, see for example %
\cite{malamud2008market} and references therein. The approaches, which are mostly of the partial equilibrium type, can be divided into four classes.
First, quite popular is the direct application of an ad-hoc chosen \emph{finance consistent (or risk-neutral)} measure, see for example \cite{BrennanSchwartz1976, Kwok2008, Krayzleretal2015, cui2017variable} and the references therein. It often leads to explicit results in a direct matter.
Second, the \emph{benchmark approach}, applied first to financial markets in  \cite{platen2006benchmark} and then to the present problem in \cite{buhlmann2003discrete}, uses the \emph{growth-optimal portfolio} as num\'eraire and evaluates risky products by expectations under $P$. For an insurance application see e.g.~\cite{biagini2015risk}.
In the \emph{local risk-minimization approach}, third, where the risk-neutral measure on the insurance side is specified by a risk-minimizing procedure (see for example \cite{Foellmer-Schweizer1989, moller2001risk, Pansera2012}) it is 
 assumed that insurance risks, like mortality risk, can be diversified away. This approach can be generalized to the quadratic hedging error. Finally, \emph{indifference pricing} leads to a non-linear pricing rule and we refer to  \cite{blanchet2015max, chevalier2016indifference} for details and further literature.
Non-linear methods to analyse insurance contracts often use the axiomatic approach to risk measures on $L^\infty$ with the Fatou property and thus admitting robust presentations, see \cite{Tsanakas2005,  BarigouChenDhaene2019, EngsnerLindensjoeLindskog2020} while this can also be treated on more general spaces, as for example in \cite{kaina2007convex, cheridito2009risk}.

Our work with the QP-rule and the BEL approach can be seen as an application of the 'two-step market evaluation' in \cite{PelsserStadje2014} in the sense that first we start by conditioning with respect to financial events (see also \cite{BarigouLIndersYang2022}). Second, we use the two same integrations with respect to conditioning and to risk neutral probability, as far as the integrand is a linear form of the contract. Of course, since we fix the statistical probability $P$, the QP-rule is \emph{time- and finance-consistent} as proposed in \cite{PelsserStadje2014}. (For time-consistency one might also have a look to  \cite{Duffie-Epstein-1992},  \cite{cheridito-etal-2006}, \cite{Delbaen-2006}, \cite{cheridito-etal-2011}, \cite{Pelsser-Ghalehjooghi-2016}). The cited authors work with nonlinear actuarial principles preventing them to obtain BEL, because of the EIOPA requirement that no prudential margin should appear in the best estimate of liabilities (BEL) (see www.actuaries.org.uk Solvency II-2016.pdf 2.2.1 and Swiss Solvency Test Technisches Dokument 3.2). As for us we follow this request and postpone the definitions of provision and solvency capital to a future paper, meaning that we start a different route beginning with the search of the best estimate. It will involve the two probability measures $P$ (historical probability) and $Q$ (useful for market consistency) defined on two different sigma-algebras the private and the public information at final date. Since long, the insurance industry has gone further, beginning with BEL (but without clear mention of the probability being used) and adding risk margin, solvency capital etc.  In complementing our work we shall introduce the objectives of shareholders, insurance seekers, and the regulator (see \cite{Albrecher-etal-2022}). From this point of view, we shall look at industry's linear best estimates, regulatory risk margin and capital requirement. 

The approach in \cite{ChenChenDhaene2020} introduces a concept of valuation, called \emph{fair}, if it is market consistent and in addition coincided with valuation under $P$ for payoffs which are independent of the future stock evolution (called $t$-orthogonal). The QP-rule  satisfies also this property.
Hedging techniques are applied to it.

In \cite{deelstra2018valuation}, a \emph{three-step method} for the valuation of hybrid insurance products is proposed consisting at first of hedging the inherent financial risk and of diversification via pooling, and for the leftover to apply a non-linear premium principle. The authors use an enlargement of filtration approach with conditionally independent extensions, thus satisfying the immersion property (see \cite{blanchet-jeanblanc2020}) . 

Some insurance products we have in mind are \emph{variable annuities}, which explicitly link the insurance benefits to the performance of financial markets. We refer to  \cite{Bacinello2011} for an extensive overview of related literature.

The paper is organized as follows: Section \ref{sec-I+F} presents the economic environment of an insurance company, containing the financial market, the insurance contracts and their allocation portfolios. It contains the different filtrations reflecting the different states of information an insurer is acting on.  Given the exogenous premiums of the standard contracts, the notion of an insurance-finance arbitrage is defined. The general fundamental theorem on non-insurance-finance arbitrage is presented in Section \ref{sec-NIFA}. Section  \ref{sec-valins} studies linear valuation rules for insurance claims and introduces the so-called QP-rule. This rule combines risk neutral pricing (under $Q$) with insurance evaluation (under $P$). In Section \ref{sec-QP-NIFA} we provide the second main theorem where a sufficient and a necessary condition for the insurance-finance consistency of the QP-rule are given. An example of a variable annuity contract in Section \ref{sec-annui} underpins this property.

\section{Insurance and finance}\label{sec-I+F}

Fundamental valuation principles are based on the absence of arbitrage with trading strategies as the central concept. In the case considered here,  strategies incorporate  trading on the financial market and selling of insurance contracts.

The well-known concept of a \emph{financial arbitrage} is a self-financing trading strategy leading to a risk-less profit. On the insurance side, the portfolio consists of \emph{allocations} of insurance contracts: The insurer has the possibility to sell standardized contracts for a large number of clients leading to a substantial reduction of risk. If this risk reduction --- in combination with trading on a financial market --- allows for a risk-less profit, we call the common portfolio an  \emph{insurance-finance arbitrage} (IFA). Here, we have to keep in mind that the insurer has a different status of information than the public one of the market.

On a probability space $(\Omega, \ccH,P)$ and a discrete finite time interval $\TT=\{0, 1, ...,T\}$, we assume that the publicly available information (life-tables, information on financial markets, etc.) is captured by the filtration $\FF=(\ccF_t)_{t\in\TT}$.
The insurance company has additional internal information (e.g.~survival times or health states concerning the population of possible clients together with historical information). This information is encoded in the filtration  $\HH=(\ccH_t)_{t\in\TT}$, which encompasses public information, i.e.
\begin{align*}
 \ccF_t \subset\ccH_t \quad \text{for } t=0,\dots,T.
\end{align*}
For our model description we start reminding the well-known concept of a financial market.

\subsection{The financial market}\label{ssec-finmar}
The financial market consists of $d+1$ tradeable securities with $\FF$-adapted price process $\tilde S=(\tilde S^0, \tilde S^1,\dots, \tilde S^d)$.
Traded assets could be bonds with and without credit risk, stocks, indices, etc. The information $\FF$ will typically be strictly larger than the filtration generated by the traded assets: economic variables like employment rates or national mortality rates are examples of such publicly available information.

The num\'eraire $\tilde S^0$ with $\tilde S^0_0=1$ may be random, but is assumed to be strictly positive. Discounted price processes are denoted by $S=(S^1,\dots,S^d)$ where $S^j = \nicefrac{\tilde S^j}{\tilde S^0}$, $j=1,\dots,d$, and $S^0\equiv 1$.

An $\FF$-\emph{trading strategy} on the financial market is  a $d$-dimensional, $\FF$-adapted process $\xi=(\xi_t)_{0 \le t \le  T-1}$ with $\xi_t=(\xi\nohbr{1}_t,\dots,\xi\nohbr{d}_t)$.
Note that the insurance company has access to more information than captured by $\FF$, such that  at a later point we will consider trading strategies which are not adapted to $\FF$ but to a larger filtration, see Section \ref{ssec-trafstrat}.

For a trading strategy $\xi$ (see for example \cite{FoellmerSchied}, Proposition 5.7), its 
\emph{(discounted) value}
is given by
 $$ V^F(\xi) :=  (\xi \cdot S)_T =
  \sum_{t=0}^{T-1} \sum_{j=1}^d \xi^j_t \; \Delta S^j_{t}, $$
 with $\Delta S_{t} = S_{t+1} - S_{t}$.
We assume throughout that the financial market does not allow for  arbitrage so that the set $\ccM_{e,b}(S,\FF)$ of equivalent martingale (or market-consistent) measures with bounded densities under which the process $S$ is a martingale on the filtration $\FF$, is not empty:
\begin{align}\label{de-NFAZ}
  \ccM_{e,b}(S,\FF)\neq \emptyset.
    \end{align}

\subsection{The insurer's standard contracts}\label{ssec-stacon}
We assume that there is a single insurer who can contract with possibly infinitely many \emph{insurance seekers}.
At each date $t\in \{0,\dots,T-1\}$, the insurer may issue contracts of one standard type\footnote{The results of this paper can easily be extended to the case where contracts are offered for a finite number of different cohorts of clients. Mathematically this amounts to simply  adding an additional index for the cohorts.}, depending on $t$. Such a contract offers coverage of  a future claim  in exchange with a premium paid at initiation of the  contract. Without loss of generality, we assume that all claims are settled at the final date $T$.

The \emph{benefits} of the \emph{standard contract} issued at date $t$ are described by a $\ccH_T$-measurable non-negative random variable $X_{t,T}$ (already discounted).
 We summarize the benefits by the process $X=(X_{t,T})_{t\in \{0,\dots,T-1\}}$.

The candidate \emph{premium} of the standard contract $X_{t,T}$ issued at $t$ is denoted by $p_t(X)$ or simply $p_t$ (already discounted); it is to be paid at date $t$ and is $\ccH_t$-measurable. The value $p_t$ has to be regarded as a \emph{basic part of premium}. On top, the insurance company adds a \emph{commercial part} which contains additional costs, risk margins, etc. Risk margins are enforced by regulation and are non-linear rules.
We will treat them in future research.

For the premiums $p_t$ we assume
\begin{align}\label{cond-pt}
  p_t \in L^1_+(\ccH_t)=L^1_+(\Omega,\ccH_t,P)
\end{align}
and summarize them by the process  $p=(p_t)_{t\in \{0,\dots,T-1\}}$.

\subsection{Insurance allocations}

The classical insurance principle of diversification to substantially reduce risk  consists in the possibility of the insurer to issue individual independent contracts with clients, called \emph{insurance seekers}. Similarly we shall assume that at each date $t<T$ the insurer can contract with possibly infinitely many insurance seekers. The associated $\ccH_T$-measurable benefits of the \emph{individual contracts} are denoted by $X^1_{t,T},X^2_{t,T},\dots\; $.

Inspired by large financial markets (e.g.~\cite{Klein2000,kabanov1995large,KleinSchmidtTeichmann2015}),
we model the insurance portfolio  as a sequence  of \emph{allocations} of contracts with insurance seekers: the \emph{allocation} $\psi_t=(\psi^{i}_t)_{i \ge 1}$ \emph{at date $t$} is a  $\ccH_t$-measurable, non-negative random sequence, i.e.~  $\psi^{i}_t$ denotes the size of the contract with the $i^{th}$ insurance seeker at date $t$.
The (discounted) value of the allocation amounts to the accumulated premiums minus the benefits over all time instances $t=0,\dots,T-1$. More precisely, we refer to 
\begin{align}
        \label{VItpsi}
        V^I(\psi):= \sum_{t=0}^{T-1} \sum_{ i \ge 1} \psi^{i}_t \big( p_t - X^i_{t,T} \big)
    \end{align}
    as
the \emph{individual insurance value} and to $V^F(\xi)$ as the \emph{global financial outcome}.

Finally, an \emph{insurance portfolio strategy} is a double sequence $\psi:=(\psi^{n}_t)_{n \ge 1, 0\le t<T}$ of allocations. We impose the following \emph{admissibility condition} for a portfolio strategy:
\emph{Convergence of the insurance volume}: there exist random variables $\gamma_t \ge  0 $, $0\le t<T$ so that
\begin{align} \label{conv-psi}
            \parallel\psi^n_t\parallel:= \sum_{i \ge 1}~\psi^{n,i}_t  \rightarrow \gamma_t \quad \text{a.s. for all } t<T.
\end{align}
The precise measurability of $\gamma_t$ is explained in the next subsection.
Later, we are interested in the particular case of a bounded portfolio strategy (see Remark \ref{rem-2NIFA}):
In addition to \eqref{conv-psi} we say that the  portfolio strategy $\psi$  is \emph{bounded} if there exists $c>0$ so that
 \begin{align} \label{bdd-psi}
            \parallel \psi^n_t\parallel \le c.
        \end{align}
for all $n\ge 1$ and $0\le t < T$.
\subsection{Trading strategies of the insurer}\label{ssec-trafstrat}
We emphasize that we do not assume completeness of the financial market $(\FF,S)$ nor do we exclude (financial) arbitrages when trading would be  done with the filtration $\HH$, where in particular information of individual insurance contracts enters (see Remark \ref{rem-info}). This information would in general allow for \emph{insider-trading}, an action prohibited by law. On the other hand, it is obvious that the financial investments of an insurer depend on the global insurance-related information, which we call \emph{macro-insurance information}, to distinguish them from the individual or \emph{micro-insurance information}. The sales volumes of the contracts are of course macro-insurance data.  

To handle this \emph{insurer's dilemma} we assume that at date $t$ the trading investment of the insurer are based on the information containing $\ccF_t$ and the macro-insurance data. More precisely, we consider an intermediary filtration $\GG=(\ccG_t)_{t \in \TT}$ with
\begin{align}\label{filH}
\ccF_t \subset \ccG_t\subset \ccH_t, \qquad t \in \TT.
\end{align}

With this filtration at hand we assume that the process $\gamma$ from Equation \eqref{conv-psi} is $\GG$-adapted. Moreover, the insurer is allowed to trade on the financial market with filtration $\GG$, i.e.\ the trading strategy $\xi$ is $\GG$-adapted.
For $t=T$, we set $\ccG_T=\ccG_{T-1}\vee \ccF_T$.

\begin{remark}\label{rem-man-arb}
  In the following we do not exclude a priori that there may be arbitrage opportunities when trading with $\bbG$-adapted strategies $\xi$, i.e.  it may very well happen that
  $\ccM_{e,b}(S,\GG)=  \emptyset$.
But, since $\ccM_{e,b}(S,\FF)\neq \emptyset$, arbitrage opportunities are only possible with the additional insider knowledge on $\GG$. We remind that in many countries there exist laws against \emph{insider-trading} and in addition special announcement regulations for managers when dealing with the stocks of their proper company.

The possibility of insider-trading using information about an individual customer  is plausible. Let the portfolio contain an important personality, say customer $i=1$, and let $\tau^1$ denote the beginning of a severe illness of customer $1$, then the knowledge of $\tau^1=1$ may allow for a arbitrage possibility, e.g. by trading derivatives on a stock related to the activity of customer $1$.
\end{remark}

\subsection{From micro- to macro-insurance data}\label{ssec-mimainsur}
The basic principle of an insurance company is the possibility to contract with a large number of insurance seekers which diversifies the individual risks. For the individual insurance benefits $X^i_{t,T}$ we generalize the classical framework of actuarial mathematics to the following conditional case. It integrates in particular the future development of the financial market and thus allows hedging parts of the insurance risks by means of trading on the  financial market.  

It turns out that the $\sigma$-algebra
\begin{align}\label{de-GtT}
 \ccH_{t,T} := \ccH_t \vee \ccG_T
\end{align}
containing the insurance information up to date $t$ and all the macro-insurance data $\ccG_T$ (including the publicly available information $\ccF_T$) plays a distinctive role.
We make the following assumptions: \begin{assumption}\label{AssX1}
    For all $t \in \TT$, the standard contract $X_{t,T}\in L^2(\Omega,\ccH_T,P)$ and the individual ones $X^i_{t,T}$ satisfy
    \begin{enumerate}
        \item $  X^1_{t,T}, X_{t,T}^2,\dots \in L^2(\Omega,\ccH_T,P)$ are  $\ccH_{t,T}$-conditionally independent,
        \item $ \ E[ X^i_{t,T}|\ccH_{t,T} ] = E[X_{t,T} | \ccH_{t,T}], \ i=1,2,\dots$, and
        \item $\Var(X^i_{t,T}|\ccH_{t,T}) = \Var(X_{t,T}|\ccH_{t,T}) < \infty,  \ i = 1,2,\dots.$
    \end{enumerate}
\end{assumption}
\begin{remark}\label{rem-info}~
  \begin{enumerate} 
\item The information about the individual contracts $X^i_{t,T}$ are included in $\ccH_T$ which --- in addition to $\ccG_T$ --- must be sufficiently large to allow existence of the sequence $(X^i_{t,T})_{i\ge 1}$ of  $\ccH_{t,T}$-conditionally independent random variables.
\item Assumption \ref{AssX1} differentiates between the  information decoded by the filtration $\GG$, and the micro-insurance information $\HH$. Condition (i) allows for \emph{conditional} independence, i.e.~the incorporation of the future evolution as well in the macro-insurance information --- like the internal development of life tables, etc. --- as in the publicly available information --- like stock markets, possible successes in medicine, etc. --- contained in $\ccG_T$ together with past micro-insurance information due to $\ccH_t$. For example we cover the case of (systemic) risk of result of future published research of new drugs and the insurance company's internal studies of possible trends in customers' longevity. Remaining risks which are neither hedgeable nor diversifiable have to be covered by a non-linear risk measure (see the three-step method in \cite{deelstra2018valuation}), generally divided between the insurer and the customer (see  \cite{Eisele-Artzner-2011}). 
  \item
The square-integrability in Assumption \ref{AssX1} could be relaxed. It is used in Proposition \ref{prop-V-mean} to conduct uniform integrability of the losses to get \eqref{temp27}. In general, however, equality \eqref{temp27} will no longer hold. Intuitively, this means that the expectation of
$$ \lim_{n \to \infty} \sum_{i \ge 1} \psi^{n,i}_t X_{t,T}^i $$
under a probability with density $L\in  L^\infty(\Omega,\ccH_{t,T},P)$ can be strictly smaller than the right-hand side of \eqref{temp27}. But in \eqref{temp31} the reduced expectation still guarantees $NIFA^\infty$ in Theorem \ref{thm-FTIFA}. 
\end{enumerate}
\end{remark}

\subsection{Insurance results}\label{ssec-insres}
On the $\sigma$-algebra $\ccH_{t,T}$ the conditional expectation of the \emph{net outcome} of the standard contract $X_{t,T}$ issued at date $t<T$ is the $\ccH_{t,T}$-measurable variable
\begin{align}\label{de-Inet}
	    	Y_{t,T}:=p_t- E_P[X_{t,T}|\ccH_{t,T}].
\end{align}
The motivation to study the net outcome in Equation \eqref{de-Inet} is the following:
Assume that at date $t$ the individual contracts $(X^i_{t,T})_{i \ge 1}$ have been sold to the insurance seekers with a volume $\gamma_t \in L^\infty_+(\ccH_t)$, for example via the \emph{uniform portfolio strategy}
 \begin{align}
    \psi^n_t=\gamma_t\cdot (1/n,\ldots,1/n,0...).     \label{temp25}
\end{align}
We call the $\GG$-adapted process $\gamma=(\gamma_0,...,\gamma_{T-1})$ the sales \emph{volume process}.

By  Assumption \ref{AssX1}, Theorem 3.5 of the conditional strong law of large numbers in \cite{Majerek2005}\footnote{ A close look to the proof of Theorem 3.5 in \cite{Majerek2005} shows that one can give general conditions on the allocations $\psi^n_t$ implying \eqref{conv-g-Y}. In particular, one must exclude the domination of a single contract, meaning that $\psi^{n,i}_t\big/\parallel \psi^n_t\parallel \rightarrow 0$ for all $t$ and all $i$.} gives for such an allocation
 \begin{align}
    \lim_{n \to \infty} \sum_{i \ge 1} \psi^{n,i}_t \Big( p_t - X_{t,T}^i \Big)
        & =     \gamma_t\;  \Big( p_t - E\big[    X_{t,T}\big|\ccH_{t,T}\big]\Big)=\gamma_t\;Y_{t,T} \label{conv-g-Y}
\end{align}
and the outcome of this strategy of a large pool is given in terms of $Y_{t,T}$. The \emph{global insurance outcome} is now
 \begin{align}
    V^I(\gamma)=\sum_{0\le t<T}\gamma_t\;Y_{t,T}  \label{globout}
\end{align}
An \emph{insurance-finance strategy} is now the pair $(\psi,\xi)$ which achieve the (discounted) \emph{insurance-finance value}
\begin{align}\label{IFR}
\lim_{n \to \infty}V^I(\psi^n)+ V^F(\xi)= \sum_{0\le t<T}\Big(\gamma_t\;Y_{t,T}+\xi_t \cdot \Delta S_t\Big).
\end{align}

\subsection{Insurance-finance arbitrage}\label{ssec-infiarb}
The \emph{insurance-finance market} is given by the triplet $(X,p,S)$ which describes the following constituents of the markets: the benefits $X$, the candidate premiums $p$ and the financial securities prices $S$ with the mentioned assumptions.
\begin{definition}\label{def-IFarb}
On $(X,p,S)$
we say that an admissible insurance portfolio strategy $(\psi^n)_{n \ge 1}$, and an insurer's trading strategy $\xi$ form an \emph{insurance-finance arbitrage} (IFA), if
\begin{align}\label{con-IFarb}
   \lim_{n \to \infty} V^I(\psi^n)+ V^F(\xi) \in L_0^+ \backslash \{0\}.
    \end{align}
If the portfolio strategy $(\psi^n)_{n \ge 1}$ is bounded, we call  $((\psi^n)_{n \ge 1},\xi)$  a \emph{bounded IFA}, otherwise we speak of a \emph{general IFA}.

If there is no general IFA on the insurance-finance market, we say  no general insurance-finance arbitrage (NIFA${}^0$) holds. If there is   no bounded IFA, we say  NIFA${}^\infty$ holds.
\end{definition}

Obviously, NIFA${}^0$ implies NIFA${}^\infty$. But the converse is not true, as the following example shows:
\begin{example}\label{ex-arb-infty}
Let $\Omega=\{\omega_1,\omega_2\}\times [0,1)$ with the two $\sigma$-algebras
\begin{align*}
 \ccF_0:=\{\emptyset,\{\omega_1,\omega_2\}\}\times \ccB([0,1)) \quad \text{and} \quad   \ccF_1:=\sigma(\{\omega_1\}, \{\omega_2\})\times \ccB([0,1))
\end{align*}
where $\ccB([0,1))$ are the Borel-sets of $[0,1)$. The probability is $P(\omega\times dz)= \half \; (\delta_{\omega_1}+\delta_{\omega_2})(\omega) \cdot dz$ with the Lebesque-measure $dz$ on $[0,1)$. As the tradeable assets we take
\begin{align*}
 \Delta S^1_0(\omega,z):=\left\{
   \begin{array}{ll}
     -1, & \hbox{if} \quad \omega=\omega_1, \\
     1, & \hbox{if} \quad \omega=\omega_2,
   \end{array}
 \right. \quad\text{and}\quad
 \Delta S^2_0(\omega,z):=\left\{
   \begin{array}{ll}
     \frac{1}{1-z}, & \hbox{if} \quad \omega=\omega_1, \\
     1-\frac{1}{1-z}, & \hbox{if} \quad \omega=\omega_2,
   \end{array}
 \right. .
\end{align*}
The strategies 
\begin{align*}
\xi^1_0=\frac{1}{1-z}\quad\text{and}\quad
\xi^2_0=1
\end{align*}
yield the arbitrage result:
\begin{align*}
\xi^1_0 \cdot \Delta S^1_0\;+\; 
xi^2_0 \cdot \Delta S^2_0 = 
\left\{
   \begin{array}{ll}
     0, & \hbox{if} \quad \omega=\omega_1, \\
     1, & \hbox{if} \quad \omega=\omega_2,
   \end{array}
 \right.
\end{align*}
while it is easy to see that there is no   arbitrage with bounded strategies. But note that neither $\xi^1_0 \cdot \Delta S^1_0$ nor $\xi^2_0 \cdot \Delta S^2_0$ are integrable with respect to $P$.
\hfill $\diamond$
\end{example}

\newpage

\begin{remark}[On the necessity of considering bounded strategies]\label{rem-2NIFA}~
\begin{enumerate}
    \item 
The reason to introduce two different notions of no-arbitrage stems from the following fact: The implication that if an 'arbitrage prohibiting' probability exists, then the 'no-arbitrage' condition holds, needs an extra integrability condition of the outcome of the strategy. In our case this leads to bounded portfolio strategies.
	The classical trick to overcome this difficulty was already observed in \cite{dalang-etal1990}, Remark 3.4 - namely one can always change to a measure $P'$ by a bounded density where integrability holds. In the presence of infinitely many assets this is no longer possible  and we will exploit this fact explicitly by specifying uniform insurance trading strategies which converge to the $P$-(conditional) expectation of insurance claims. 

\item Bounded insurance strategies were first studied in \cite{CarassusPhamTouzi2001} where the set of admissible strategies forms a closed convex cone. 
\end{enumerate}
\end{remark}

In the setting considered here with two assets on different filtrations, many subtleties arise and one can not proceed as usual for the proof of a fundamental theorem. This was already observed in \cite{KabanovStricker2006, CuchieroKleinTeichmann2020}. In particular, there might be examples of two-period arbitrage.
\begin{example}\label{ex-arb-T2}
Inspired by \cite{KabanovStricker2006}, we construct a two-period model containing a random set $A$ with $P(A)\in (0,1)$ and a sequence $(A^i)_{i\ge 1}$ of conditionally independent sets satisfying
\begin{align*}
  E[\Ind_{A^i}\;|\;\sigma(A)]=\Ind_{A}.
\end{align*}
We set $\ccF_0=\ccG_0=\ccH_0=\ccF_1=\ccG_1=\{\emptyset,\Omega\}$, $\ccF_2=\ccG_2=\sigma(A)$, $\ccH_1=\sigma((A^i)_{i\ge 1})$, and $\ccH_2=\sigma(A, (A^i)_{i\ge 1})$ such that $\ccH_{0,2}=\ccF_2$.
A possible interpretation is as follows: Let $A^i$ be the outbreak of an epidemic illness of the individual customer $i$. At date $t=1$ these individual outbreaks are known to the insurer, but the general event $A$ is known publicly only at date $t=2$. 

The insurance contracts are
\begin{align*}
 X^i_{0,2}=3/2 \; \Ind_{A^i} \quad \text{and} \quad   X^i_{1,2}=0,
\end{align*}
together with the standard strategy $\psi^n_0=(1/n,\ldots,1/n,0...)$ and a premium $p_0=1$ so that
\begin{align*}
 Y_{0,2}=p_0-E[X^i_{0,2} \;|\; \ccH_{0,2}]=\Ind_{A^c}- 1/2\;\Ind_{A}.
\end{align*}
For the financial market we take
\begin{align*}
& S_0=S_1=1 \quad \text{and} \quad S_2=2\; \Ind_{A}+ 1/2\;\Ind_{A^c}&\text{so that}\\
&\Delta S_1=\Ind_{A} - 1/2\;\Ind_{A^c}.
\end{align*}
With the financial strategies $\xi_0=0$ and $\xi_1=1$ we find that
\begin{align*}
  \lim_{n \to \infty} V^I(\psi^n)+ V^F(\xi)=1/2,
\end{align*}
an insurance-finance arbitrage.
\hfill $\diamond$
\end{example}

\section{The  fundamental theorem of insurance-finance arbitrage}
\label{sec-NIFA}
In this section we analyse the insurance-finance market and characterize when NIFA holds.
Combining trading on the financial market with the net outcome on the insurance markets yields the following set of possible terminal values generated from insurance-finance trading, described by the cone $\ccK$:
\begin{align}\label{de-K}
	    	&\ccK:=\Big\{\sum_{0\le t<T}\big(\gamma_t \, Y_{t,T} + \xi_t \, \Delta S_t\big) \,\big|\, \gamma_t\in L^0_+(\ccG_t,\RR^1), \, \text{ and}\; \xi_t\in L^0(\ccG_t,\RR^d)\Big\}.
\end{align}
We note that all random variables in $\ccK$ are $\ccH_{T-1,T}$-measurable.

For the following theorem we will also rely on  a measure $P^*$ under which we will assume that Assumption \ref{AssX1} is satisfied. In addition we will assume that the conditional expectation of $X_{t,T}$ under $P^*$ coincides with those under $P$: 

\begin{assumption}\label{AssX1*}
Consider $P^* \sim P$ and assume that 
   for all $t \in \TT$, 
    \begin{enumerate}
        \item $  X^1_{t,T}, X_{t,T}^2,\dots \in L^2(\Omega,\ccH_T,P^*)$ are  $\ccH_{t,T}$-conditionally independent under $P^*$,
        \item $ \ E_{P^*}[ X^i_{t,T}|\ccH_{t,T} ] = E_{P^*}[X^1_{t,T} | \ccH_{t,T}], \ i=2,3,\dots$, and
        \item $\Var_{P^*}(X^i_{t,T}|\ccH_{t,T}) = \Var_{P^*}(X^1_{t,T}|\ccH_{t,T})< \infty,  \ i = 2,3,\dots.$
    \end{enumerate}
\end{assumption}

\renewcommand{\theenumi}{\roman{enumi}}
\renewcommand{\labelenumii}{(\theenumi.\theenumii)}
\makeatletter\renewcommand{\p@enumii}{\theenumi.}
\makeatother
We can now formulate our main result:
\begin{theorem}\label{thm-FTIFA}
On the insurance-finance market $(X,p,S)$ with Assumption \ref{AssX1},  the sequence of implications
 \eqref{FTIFA1}$\Rightarrow$
\eqref{FTIFA2}$\Rightarrow$\eqref{FTIFA3} \; holds for the following assertions:
  \begin{enumerate}
    \item\label{FTIFA1}\qquad $NIFA^0$ holds,
    \item\label{FTIFA2}\qquad
$\big(\ccK-L^0_+(\ccH_{T-1,T})\big) \cap L^0_+(\ccH_{T-1,T})=\{0\}$,
    \item\label{FTIFA3}\qquad There exists  $P^*\sim P$ on $(\Omega,\ccH_{T-1,T})$ so that
\begin{enumerate}
  \item\label{coFT31}
  $\; P^*|_{ \ccG_T} \in \ccM_{e,b}(S,\GG)\;$ and
  \medskip
  \item\label{coFT32}$\; E_{P^*}\big[Y_{t,T}\big|\ccG_t\big] \le 0$ for $t=0,\dots,T-1$.
\end{enumerate}
  \end{enumerate}
  Moreover, if \eqref{FTIFA3} holds and $P^*$ satisfies Assumption \ref{AssX1*}, then 
 \begin{enumerate}
 	\setcounter{enumi}{3}
 	\item  \label{FTIFA4}\qquad $NIFA^\infty$ holds.\end{enumerate}
\end{theorem}
\medskip
\begin{proof}We begin by showing \eqref{FTIFA1}$\Rightarrow$\eqref{FTIFA2}.
Assume that we have
 \begin{align}
  \gamma\cdot Y + \xi \cdot \Delta S=\sum_{t=0}^{T-1}\gamma_t\;  \Big( p_t - E\big[    X_{t,T}\big|\ccH_{t,T}\big]\Big)+ \xi \cdot \Delta S\in L^0_+(\ccH_{T-1,T})\setminus \{0\}.
   \label{temp12}
\end{align}
We first consider the uniform portfolio strategy with the volume $\gamma_t$:
 \begin{align*}
    \psi^n_t=\gamma_t\cdot (1/n,\ldots,1/n,0...).
\end{align*}
By  Assumption \ref{AssX1}, Theorem 3.5 of the conditional strong law of large numbers in \cite{Majerek2005} gives for such an allocation the following result:
 \begin{align*}
     \lim_{n \to \infty} \sum_{i \ge 1} \psi^{n,i}_t \Big( p_t - X_{t,T}^i \Big)
        & =     \gamma_t\;  \Big( p_t - E\big[    X_{t,T}\big|\ccH_{t,T}\big]\Big)=:\,\gamma_t\;Y_{t,T}.
\end{align*}
Therefore \eqref{temp12} yields
 \begin{align*}
  \lim_{n \to \infty} V^I(\psi\nohbr{n})+ V^F(\xi)
  =\gamma_t\;Y_{t,T}+ \xi \cdot \Delta S\in L^0_+(\ccH_{T-1,T})\setminus \{0\}
\end{align*}
a contradiction to $NIFA^0$. This shows $\eqref{FTIFA1}\Rightarrow\eqref{FTIFA2}$
\medskip

Next, we show  $\eqref{FTIFA2} \Rightarrow \eqref{FTIFA3}$. 
Changing $P$ by a suitable bounded strictly positive density to some $P_1$ allows to assume $Y_{t,T}$ and $\Delta S_t$ in $L^1(\ccH_T)$. By Proposition \ref{aux3} also $\ccK \cap L^1(\ccH_{T-1,T})-L^1_+(\ccH_{T-1,T})$ is closed in $L^1(\ccH_{T-1,T})$. The Kreps-Yan Theorem gives another bounded strictly positive density, leading to the probability $P^*$ so that $E_{P^*}[V_0] \le 0$ for all $V_0 \in \ccK$. This implies
$\pm E_{P^*}[\Ind_{A_t}\cdot \Delta S_t] \le 0$ for all $A_t\in \ccG_t$ so that $P^*|_{ \ccG_T} \in \ccM_{e,b}(S,\GG)$ which shows \eqref{coFT31}.
Similarly we get  $E_{P^*}\big[\Ind_{A_t}\;Y_{t,T}\big] \le 0$ which shows \eqref{coFT32} of the Theorem.
\medskip

With  Proposition \ref{prop-V-mean*} we prove the final implication of Theorem \ref{thm-FTIFA}.
Let $P^*$ be an equivalent probability measure on $(\Omega,\ccH_{T-1,T})$ with density $L=dP^*/dP $ bounded by $C$ and satisfying \eqref{coFT31} and \eqref{coFT32} Further, assume that we had a bounded insurance-finance arbitrage, i.e.
\begin{align}\label{ifarbit}
   \lim_{n \to \infty} V^I(\psi\nohbr{n})+ V^F(\xi) \in L_0^+ \backslash \{0\}
\end{align}
for some bounded portfolio strategy $(\psi^n)_{n\ge 1}=(\psi^n_t)_{n\ge 1, t<T}$ and some financial strategy $(\xi_t)_{t\le T}$.

First, $\lim_{n \to \infty} V^I(\psi\nohbr{n})+ V^F(\xi) \ge 0$ implies with \eqref{bdd-psi} that
\begin{align*}
V^F(\xi) \ge - \lim_{n \to \infty} V^I(\psi\nohbr{n}) \ge - \sum_{t=0}^{T-1}\lim_{n \to \infty}\sum_{i \ge 1}\psi_t^{n,i} p_t \ge - c \sum_{t=0}^{T-1} p_t.
\end{align*}
Since \eqref{cond-pt} implies  that  $E_{P^*}[p_t]=E_P[L \cdot p_t] \le C\, E_P[p_t]< \infty$, for $t=0,\dots,T-1$, it follows that $E_{P^*}[V_T(\xi)^-]< \infty$ and we obtain
\begin{align}\label{temp11}
    E_{P^*}[ V^F(\xi)]=0.
\end{align}

This property yields
\begin{align} \label{temp31}
E_{P^*}\Big[  \lim_{n \to \infty} V^I(\psi\nohbr{n})+ V^F(\xi)   \Big]
& =  E_{P^*}\Big[  \lim_{n \to \infty} V^I(\psi\nohbr{n}) \Big] \notag \\
&= \sum_{t < T}  E_{P^*}\Big[  \lim_{n \to \infty}\sum_{i \ge 1} \psi^{n,i}_t \Big( p_t - X_{t,T}^i \Big) \Big]. 
\end{align}
By Equation \eqref{conv-psi} together with Proposition  \ref{prop-V-mean*},
\begin{align}
	\eqref{temp31} &= \sum_{t < T}  E_{P^*}\big[  \gamma_t p_t \big] -  E_{P^*}\Big[  \lim_{n \to \infty}\sum_{i \ge 1} \psi^{n,i}_t  X_{t,T}^i  \Big] \notag \\
	&= \sum_{t < T}  E_{P^*}\big[  \gamma_t p_t \big] -  E_{P^*}\Big[ \, \gamma_t    E_P\big[X_{t,T}\big|\ccH_{t,T}\big]\Big].
\end{align}
Property \eqref{coFT32} of $P^*$ and the non-negativity of $\gamma_t$, yields
\begin{align*}
E_{P^*}\Big[ \gamma_t \, \big(p_t - E_P[ X_{t,T}|\ccH_{t,T}]\big)\Big]
 & = E_{P^*}\Big[ \gamma_t \, E_{P^*}\big[Y_{t,T}\big|\ccG_t\big]\Big]\le 0.
\end{align*}
Hence, Equation \eqref{temp31} is non-positive, which is a contradiction to the assumption of an insurance-finance arbitrage.
The part $\eqref{FTIFA3}\Rightarrow\eqref{FTIFA4}$ of Theorem \ref{thm-FTIFA} is proven, too.
\end{proof}
\begin{remark}\label{cond-integr}
    The work \cite{KabanovStricker2006} is the one which is most related to our work. The authors consider trading strategies which are adapted to a reference filtration, but stock prices do not need to be adapted. 
    In the insurance-finance case presented here, the probability $P^*$ is characterized by two separate properties: \eqref{coFT31} and \eqref{coFT32} corresponding to the two elements $V^I$ and $V^F$ of the insurance-finance result. Therefore the additional boundedness condition \eqref{bdd-psi} is needed.   
\end{remark}

\begin{remark}[On the necessity of Assumption \ref{AssX1*}] Condition \eqref{FTIFA3} is not enough to imply the absence of insurance arbitrage: indeed, this is visible in \eqref{temp31}, where the term
\begin{align} 
 E_{P^*}\Big[  \lim_{n \to \infty}\sum_{i \ge 1} \psi^{n,i}_t \Big( p_t - X_{t,T}^i \Big) \Big] 
\end{align}
is considered. 
Passing from $X^i$ to $X$ is possible under $P$ by Assumption \ref{AssX1}, but with an absolutely continuous change of measure the expectation under $P$ may be  greater than zero. Assumption \ref{AssX1*} allows to apply Proposition \ref{prop-V-mean*} and hence puts us in the position to apply property \eqref{coFT32}. 
\end{remark}

\subsection{Market consistency}
In mathematical finance the notion of a \emph{market-consistent evaluation} was, to the best of our knowledge, first introduced by \cite{Cont2006}, \cite{cheridito2008dynamic}, Malamud et al. (2008), \cite{artzner2010supervisory}. See also \cite{PelsserStadje2014}, among others. This property is mainly applied to non-linear evaluations (like risk measures) of risky situations, requiring that the evaluation acts linearly on traded positions via their market values. In our context it would be preferable to call this property \emph{finance-consistent}. In the linear case\footnote{The property of insurance-finance-consistent evaluations plays also an important role for non-linear operators, like risk measures. They appear at the calculation of risk margins (RM) and solvency capital requirements (SCR) for insurance companies (see \cite{Hainaut-etal-2018}). In particular, the calculation of the risk margin should be market consistent, since in bookkeeping it is classified as external capital and therefore should not include risk which can be hedged by the insurance company.}, it can be naturally extended to the following notion:

\begin{definition}
An equivalent probability measure $P^*$ on $(\Omega,\ccG_{T-1,T})$ is called \emph{insuran\-ce-finance-consis\-tent} if it satisfies the conditions \eqref{coFT31} and \eqref{coFT32} of Theorem \ref{thm-FTIFA}.
\end{definition}
Such a consistent probability measure can be used for arbitrage-free valuation in the usual sense: indeed, if $P^*$ is insurance-finance consistent and we consider the so-called \emph{reference premium}, associated to $P^*$, 
\begin{align}\label{p-star}p^*_t = E_{P^*}[X_{t,T} | \ccH_t],
\end{align}
then conditions \eqref{coFT31} and \eqref{coFT32} of Theorem \ref{thm-FTIFA} are met and --- given that $P^*$ satisfies Condition \ref{AssX1*} --- there is no insurance-finance arbitrage with respect to the reference premium $p^*$ in the sense of NIFA${}^\infty$.

We study a case on a  finite probability space in the following example.

\begin{example}\label{ex-FTIFA1}
We consider two points in time,  $\TT=\{0,1\}$, and  $\Omega=\{\omega_1,\omega_2,\omega_3,
\omega_4\}$ with the two filtrations
$\FF=(\ccF_0,\ccF_1)$ and $\GG=(\ccG_0,\ccG_1)$ given by
\begin{align*}
    \ccF_0&=\{\emptyset,\Omega\}, \; &\ccF_1&=\sigma(\{\omega_1,\omega_3\},\{\omega_2,
\omega_4\})&\qquad\text{resp.}\\
    \ccG_0&=\sigma(\{\omega_1,\omega_2\},\{\omega_3,\omega_4\}), &\ccG_1&=\ccP\big(\Omega\big).&
\end{align*}
Besides the constant asset $S^0\equiv 1$, the risky asset of a financial market is $S^1_0\equiv 1$ and $S^1_1=\threehalf\ind{\omega_1,\omega_3}
+\half\ind{\omega_2,\omega_4}$ such that on the filtration $\FF$ we have a complete market with the unique finance-consistent probability:
\begin{align*}
Q(\{\omega_1,\omega_3\})=Q(\{\omega_2,\omega_4\})=\half.
\end{align*}
The statistical probability $P$ is supposed to be
\begin{align*}
P(\{\omega\})=0.3\,\ind{\omega_1,\omega_2}
+0.2\,\ind{\omega_3,\omega_4}
\end{align*}
and as such it is finance-consistent since $P(\{\omega_1,\omega_3\})=P(\{\omega_2,\omega_4\})=\half$, i.e.~$P|_{\ccF_T}=Q$.

For the standard insurance contract, we assume
\begin{align*}
p_0&=4\,\ind{\omega_1,\omega_2}+2\,\ind{\omega_3,\omega_4} \\
X_{0,1}&=5\,\ind{\omega_1}+2\,\ind{\omega_2}
+3\,\ind{\omega_4}\qquad\text{such that the net outcome is}\\
Y_{0,1}&=p_0-X_{0,1}=-1\,\ind{\omega_1}+2\,\ind{\omega_2}
+2\,\ind{\omega_3}-\ind{\omega_4}
\end{align*}
Since also any insurance-finance-consistent probability $P^*$ must be finance-consistent, it is of the form
\begin{align*}
P^*=\half\,\big( \alpha \ind{\omega_1}+\beta    \ind{\omega_2}+\,(1-\alpha)\ind{\omega_3}
+\,(1-\beta)\ind{\omega_4}\big).
\end{align*}
with $\alpha,\beta \in (0,1)$. This implies that
\begin{align*}
E_{P^*}\big[Y_{0,1}\big|\ccG_0]=
\left\{\begin{array}{ll}
  \frac{2\beta-\alpha}{\alpha+\beta}, & \hbox{on the set}\;\{\omega_1,\omega_2\} \\
  \frac{1+\beta-2\,\alpha}{2-(\alpha+\beta)}, & \hbox{on the set}\;\{\omega_3,\omega_4\}.\end{array}\right.
\end{align*}
Hence, $P^*$ is equivalent and insurance-finance-consistent if and only if \begin{align*}
\alpha > \half \qquad\text{and }\: 0<\beta \le \min(\alpha/2, 2\,\alpha-1).
\end{align*}
Therefore, here the statistical probability $P$ is finance-consistent but \textbf{not} insurance-finance-consistent.

On the other hand, if instead of $p_0$ the premium is raised to $\widetilde{p}_0=5\,\ind{\omega_1,\omega_2}+3
\,\ind{\omega_3,\omega_4}$, then obviously we have an insurance-finance arbitrage.
\hfill $\diamond$
\end{example}

\section{Valuation of non-traded wealth: the QP-rule and EIOPA}
\label{sec-valins}
Insurance companies have a sound collection of statistical data out of which they create a sufficiently reliable probability $P$ for the events they are interested in. On the other hand, insurance contracts are more and more linked to financial markets via interest rates, variable annuities, equity-linked life insurance etc., so-called hybrid products. These contracts, once sold, are not traded on markets. Therefore often the statistical valuation and the market valuation of the contracts do not coincide. By the fundamental theorem of asset pricing  it is commonly agreed upon that market valuation should be done by a \emph{risk neutral} --- better called \emph{finance-consistent} --- probability measure (cf. Theorem \ref{thm-FTIFA}). Hence a consistent valuation rule linking traded assets with non-traded insurance contracts is of highest importance.
Moreover, as already mentioned before, insurance companies have more information than the publicly available information of financial markets. 

Altogether we therefore face two measures, the statistical measure $P$ and a risk neutral one,
defined on different $\sigma$-fields and are seeking a solution of  the following problem: 
given two $\sigma$-algebras $\ccF \subset \ccH$ on a set $\Omega$, a probability $Q$ on $(\Omega,\ccF)$, and a probability $P$ on $(\Omega,\ccH)$, how should we extend $Q$ to $(\Omega,\ccH)$ in a $P$-reasonable way. We assume that $Q$ and the restriction $P\big|\ccF$ of $P$ to $\ccF$ are equivalent.

We record a key concept in the area, the $\QcirP$ probability measure, which exploit the two different properties of $P$ and $Q,$ namely the statistical information and the financial market information.  
The first to propagate this tool were \cite{plachky1984conservation}, studying measure extensions in a statistical context. The same problem arises in the context of time-consistent dynamic risk measures, see for example \cite{cheridito-etal-2006}.
Already \cite{dybvig1992hedging} proposed to use this tool for the evaluation of non-traded wealth in the framework of state-price densities.  
\cite{PelsserStadje2014} use this method to build a coherent risk measure with a leading probability term $Q$ in the context of actuarial premium principles.

We start by formulating the measure-extension from \cite{plachky1984conservation} in a context with two filtrations. We work on a measurable space $(\Omega,\ccH)$ with  two filtrations $\FF=(\ccF_t)_{t \in \TT}$ and $\HH=(\ccH_t)_{t \in \TT}$ such that $\ccF_t \subset \ccH_t$ for all $t \in \TT$. Note that the result can be extended to  $T=\infty$ under the condition that there exists a martingale measure $\Q$ on $(\Omega,\ccF_\infty)$. For simplicity we write $\ccF$ and $\ccH$ instead of $\ccF_T$, resp. $\ccH_T$ in the following.

\begin{proposition}\label{prop:PQ-rule}
There is a unique probability measure, denoted by $\QcirP$, on $(\Omega,\ccH)$ such that $\QcirP = Q$ on $\ccF$ and for all $A \in \ccH$ it holds that $ \QcirP(A \,|\, \ccF ) = P ( A \,|\,\ccF ) $. This measure $\QcirP$ satisfies for any random variable $X \ge 0$ that
\begin{enumerate}
    \item\label{prop:PQ-rule:i} for a $\sigma$-algebra $\ccG\subset \ccF$, 
\begin{align}\label{fo-QP1}
    E_{\subQcirP}[X\,|\,\ccG]  & = E_Q \big[ E_P [ X \,|\,\ccF ] \,|\,\ccG\big],
\end{align}
    \item \label{prop:PQ-rule:ii} for a $\sigma$-algebra $\ccG$ satisfying $\ccF\subset \ccG \subset \ccH$, 
\begin{align}\label{fo-QP2}
    E_{\subQcirP}[X\,|\,\ccG]  & =
    E_P \big[ X \, |\,\ccG\big].
\end{align}
\end{enumerate}
\end{proposition}

Intuitively, the result states that $\QcirP$ coincides with the pricing measure $Q$ on the information of the financial market, thus is market-consistent. From \eqref{prop:PQ-rule:i}, the measure $\QcirP$ provides: for $t \in \TT$,
and for $X \in L^1$ or bounded from below,
        \begin{align}\label{PQ-rule:filtation}
          E_{\subQcirP}[X \,|\, \ccF_t]  & = E_Q \big[ E_P [ X \,|\,\ccF_T ]\, |\,\ccF_t \big].
    \end{align}

\begin{proof}
We construct the measure $\QcirP$ as follows: let $L$ denote the Radon-Nikodym derivative $L$ of $Q$ with respect to $\P|_{\ccF}$, such that $dQ=L\; dP$. Define $\QcirP$ on $(\Omega, \ccH)$ by
\begin{align} \label{def:QcP}
            d(\QcirP) = L\, dP.
\end{align}
such that $\QcirP(A) = E_P[ \Ind_A L ]$ for $A\in \ccH$. Obviously $\QcirP$ and $Q$ coincide on $\ccF$ by construction. Moreover, for $A \in \ccH$ we have that
\begin{align*}
    \int_B \Ind_A\; d(\QcirP) &=
    \int_B \Ind_A L\; dP = \int_B L \;P(A\,|\,\ccF)\; dP  \notag \\
    & = \int_B P(A \,|\,\ccF)\; d(\QcirP), \quad \text{for all}\;B \in \ccF,
\end{align*}
such that  $ \QcirP(A \,|\, \ccF ) = \P ( A \,|\,\ccF ) $.\\
For uniqueness, consider a further measure $R$, such that $R=Q$ on $\ccF$ and such that $ R(\cdot | \ccF ) = \P ( \cdot |\ccF ) $. Then, for $A \in \ccH$ it holds that
     \begin{align*}
        R(A) = \int R(A | \ccF )\; dR = \int \P(A|\ccF)\; dR = \int  \P(A|\ccF)\; dQ  = \QcirP(A)
     \end{align*}
     and we obtain that $\QcirP=R$.\\
In view of \eqref{prop:PQ-rule:i}, we obtain with $A\in \ccH$ that
\begin{align*}
     \int_B\;\Ind_A\;d(\QcirP)  =
     \int_B\;L\;P(A \,|\,\ccF)\; dP =
     \int_B\;P(A \,|\,\ccF)\; dQ \quad \text{for all}\;B \in \ccG\subset\ccF,
\end{align*}
so that $E_{\subQcirP}(\Ind_A \,|\, \ccG)=E_Q(P(A \,|\,\ccF)\,|\,\ccG)$ and an application of the monotone class theorem yields \eqref{prop:PQ-rule:i}.\\
Finally, for \eqref{prop:PQ-rule:ii} we observe that by Bayes' rule, it follows from $\ccF\subset \ccG \subset \ccH$ that
\begin{align*}
    E_{\subQcirP}[X\,|\,\ccG] = \frac{E_P[L \;X \,|\,\ccG]}{ E_P[L\,|\,\ccG]}=\frac{L \;E_P[X \,|\,\ccG]}{ L} = E_P[X \,|\, \ccG],
\end{align*}
since $L$ is strictly positive a.s. and $\ccF$-measurable and hence $\ccG$-measurable.
\end{proof}

Under our assumption \eqref{de-NFAZ}
in Section \ref{ssec-finmar} that the financial market $(S,\FF)$ is free of arbitrage with respect to publicly available information $\FF$, we have a finance-consistent measure $Q\in \ccM_{e,b}(S,\FF)$ on $(\Omega,\FF)$. We emphasize that for the formation of $Q$ only trading strategies adapted to $\FF$ are used. Applying Proposition \ref{prop:PQ-rule} to general values, we get a tool --- called  \emph{QP-rule}\footnote{The probability $\QcirP$ is also called \emph{pasting} of $Q$ with $P$.} --- to evaluate in the spirit of \cite{dybvig1992hedging} non-traded wealth in a way excluding financial arbitrage.\\ 
In combination with the probability $P$ statistically gained by the insurer, the QP-rule yields a \emph{reference premium} $p^*$ associated to $\QcirP$ for benefits $Y$:
\begin{align}\label{ref-premQP}
    p^* = E_{\subQcirP}[Y].
\end{align}
In this case, there is no insurance-finance arbitrage if the premium $p$ asked for the benefit $Y$ is less than $p^*$. 
\medskip

In our opinion
it is also to be used for computing the regulatory best estimate of liabilities (BEL)  required in the European Directive 2009/1388/CE Article 77: \emph{The best estimate shall correspond to the probability-weighted average of future cash-flows taking account of the time value of money (expected present value of future cash-flows using the relevant risk-free interest rate term structure)}. Moreover the fact that the QP-rule  is linear does agree with the regulation that no prudential margin should appear in the BEL.

\begin{example}[A discrete example of an hybrid insurance contract]\label{exa-discr}
Consider a one-period model with  three states on the financial market: high, middle, and low  ($h,m,l$).
We assume two cases on the insurance side: insured situation and not insured situation ($in, out$). This corresponds to
$$ \Omega = \big\{(h,in),(h,out),(m,in),(m,out),(l,in),(l, out)\big\}.$$
Denote the high, middle and low states of the market by $h=\{(h,\, in),(h,\, out)\}$, $m=\{(m,\, in),$ $(m,\, out)\}$, and $l=\{(l,\, in),(l,\, out)\}$. Similar for the insurance cases: $in=\{(h,\, in),(m,\, in), (l,\, in)\}$ and  $out=\{(h,\, out),(m,\, out), (l,\, out)\}$.
The information on the financial market is given by $\ccF = \sigma \big(h,\, m,\, l\big)$, while $\ccG=\ccH= \ccP(\Omega)$.
Besides the constant equal to $1$  num\'eraire, there is a risky asset $S$ with the values $S_0=1$  and $S_1(h)=1.5$, $S_1(m)=1$, and $S_1(l)=0.5$.
Below, we shall discuss different cases of probabilities $P$ on $\ccH$.

We assume that the insurance offers a hybrid contract of the type:
$$X_{0,1}=\ind{in}\cdot \big(S_1-0.7\big)^+,$$
i.e. a call in the risky asset with strike price 0.7 to be paid only in the insured situation $in$. This contract is offered at a premium $p_0$ and sold by a uniform portfolio strategy $\psi^n_t=(1/n,\ldots,1/n,0...)$. Using a financial strategy $\xi_0$ for the risky asset $S$, the insurance-finance result is
$$\lim_{n \to \infty} V^I_{T}(\psi\nohbr{n})+ V^F_{T}(\xi) = p_0- \ind{in}\cdot \big(S_1-0.7\big)^+ \;+ \; \xi_0\cdot \big(S_1-1\big).$$
We want to apply the QP-rule to this contract and investigate when it is insurance-finance-consistent.

First, we start with a \emph{complete market} by assuming that $P$ is given by the vector on $\Omega$:
$$\big(0.1\;\alpha, 0.9\;\alpha,\; 0,\; 0 ,\;0.4\;(1-\alpha),\;0.6\;(1-\alpha)\big).$$
with $\alpha \in (0,\, 1)$. The unique finance-consistent probability on $\ccF$ is
$Q=\half \big( \ind{h}+ \ind{l}\big)$. As the conditional probability we find $$P(\cdot|\ccF)= \ind{h}\;\big(0.1, \,0.9, \,0, \,0, \,0, \,0\big)+\ind{l}\;\big(0, \,0, \,0, \,0, \,0.4, \,0.6\big),$$
independent of $\alpha$. For the QP-rule we get
$\QcirP(\cdot)=\big(0.05, \,0.45, \,0, \,0, \,0.2, \,0.3\big).$

Hence, $$E_{\subQcirP} \Big[\lim_{n \to \infty} V^I_{T}(\psi\nohbr{n})
+ V^F_{T}(\xi)\Big] = p_0-0.8*0.05=p_0-0.04.$$
The probability $\QcirP$ is insurance-finance-consistent
if and only if $p_0 \le 0.04$. The risk minimizing financial strategy (in the sense of $\essinf$) is $\xi^*_0=0.8$.

Second, we consider an \emph{incomplete market} where the physical probability $P$ has the vector form
$$\big(0.1\;\alpha, 0.9\;\alpha,\; 0.2\;\beta,\; \; 0.8\;\beta, ,\;0.4\;(1-\alpha-\beta),\;0.6\;(1-\alpha-\beta)\big).$$
with $\alpha, \beta, 1-\alpha-\beta \in (0,\, 1)$.
The set of finance-consistent measures on $\ccF$ is given by $Q^\eta(h)=\eta$, $Q^\eta(m)=1-2\,\eta$, and $Q^\eta(l)=\eta$, with $\eta \in (0,\half)$. The conditional probability $P(\cdot|\ccF)$ is now
$$P(\cdot|\ccF)= \ind{h}\;\big(0.1, \,0.9, \,0, \,0, \,0, \,0\big)+\ind{m}\;\big(0, \,0, \,0.2, \,0.8, \,0, \,0\big)+\ind{l}\;\big(0, \,0, \,0, \,0, \,0.4, \,0.6\big)$$
and $Q^\eta\mkern-2mu\odot\mkern-2mu P$ has the vector form
$\big(0.1\,\eta, \,0.9\,\eta, \,0.2\,(1-2\,\eta), \,0.8\,(1-2\,\eta), \,0.4\,\eta, \,0.6\,\eta\big).$ It follows that
\begin{align*}
        E_{\subQetacirP} \Big[\lim_{n \to \infty} V^I_{T}(\psi\nohbr{n})
+ V^F_{T}(\xi)\Big] = p_0-0.06+0.04 \eta.
    \end{align*}
Thus, if $p_0\le 0.04$, then all QP-rules of the form $\Q^\eta\mkern-2mu\odot\mkern-2mu\P$ are insurance-finance-consistent.
Again, the risk minimizing financial strategy is $\xi^*_0=0.8$, and only if $p_0>0.4$ there is the possibility of insurance-finance arbitrage. This shows that even if all QP-measures are insurance-finance-consistent, it does not exhaust all cases of non-insurance-finance arbitrages.\\
The reader might have noticed that we are able to deal with dependence of financial and insurance risks from the very beginning by not imposing a product structure to the probability $P$, in particular we should be better prepared for the cases of hybrid products which do not allow to fully disentangle the two risks.
\hfill $\diamond$
\end{example}

\section{No-arbitrage bounds under insurance-finance consistency} \label{sec-QP-NIFA}

The classification of absence of arbitrage in financial markets often utilizes the well-known no-arbitrage price bounds, given by suprema and infima over prices computed under equivalent martingale measures.
In this section we will establish similar tools in the insurance-finance context. 

Recall that 
$$ \ccF_t\;\subset\;\ccG_t\;\subset\;\ccH_t \qquad\text{for all}\; t\in \TT.$$
We define 
for a  random variable $\zeta$ and a $\sigma$-field $\ccF \subset \ccH$ \footnote{Compare Proposition 2.6 in \cite{BarronCardaliagutJensen2003}. }
    \begin{align}
    	\essinf_{\ccF} \zeta  := \esssup\, \{ \eta \in L^0(\ccF) \,| \,  \eta \le \xi \}
    \end{align}
and similarly for a set of random variables instead of $\zeta$. 

The following notions reveal to be important. 
Let
\begin{align}  \label{de-delta-t}
    \delta_{t,T} :=\essinf_{\ccF_T} \,\big(p_t- E_P[X_{t,T}\,|\,\ccH_{t,T}]\big)
\end{align}
and for $t<T$
\begin{align}  \label{pit-up-down}
    \pi^\uparrow_t(\delta_{t,T}):=\underset{\scriptscriptstyle{Q \in \ccM_{e,b}(S,\FF)}}{\esssup}\; E_Q[\delta_{t,T}\,|\,\ccF_t]\qquad \text{ and }\qquad \pi^\downarrow_t(\delta_{t,T}) =\underset{\scriptscriptstyle{Q \in \ccM_{e,b}(S,\FF)}}{\essinf}\; E_Q[\delta_{t,T}\,|\,\ccF_t].
\end{align}

Our second result is the following theorem about the absence of insurance-finance consistency.
  \begin{theorem} \label{thm-NIFA}
 Consider the insurance-finance market  $(X,p,S)$ with Assumptions \ref{AssX1}.
If there exists $t<T$, so that $\delta_{t,T}$ is bounded from above and
 \begin{align}  \label{p-cond-B}
    P\big(\pi^\uparrow_t (\delta_{t,T})> 0\quad \text{and} \quad \pi^\downarrow_t(\delta_{t,T}) \ge 0\big)>0
\end{align}
then there exists an insurance-finance arbitrage (IFA) with bounded portfolio strategy and the $\QcirP$ measure is not insurance-finance consistent. 
\end{theorem}

\begin{proof} 
Assume that for one $t<T$ we have a set
$A_t:= \big\{\pi^\uparrow_t > 0\: \text{and}\: \pi^\downarrow_t \ge 0\big\}\in \ccF_t$ with $P(A_t)>0$.
Since $\Ind_{A_t}\delta_{t,T}$ is bounded from above and $\Ind_{A_t}\pi^\downarrow_t(\delta_{t,T})$ is the conditional sub-hedging price of $\Ind_{A_t}\delta_{t,T}$, Proposition 3.14 in \cite{NiemannSchmidt2021} yields that there is a sub-hedging strategy $\xi=\Ind_{A_t}\,\xi$ so that
\begin{align} \label{temp28}
     \Ind_{A_t}\delta_{t,T}+	  \sum_{s=t}^{T-1} \xi_s \cdot \Delta S_s  &\ge \Ind_{A_t}\pi^\downarrow_t \ge 0.
\end{align}
since
\begin{align*}
&\text{on the set}\; A_t\cap \{\pi^\uparrow_t =\pi^\downarrow_t\}\; \text{we have }\;
&& \delta_{t,T}+ \sum_{s=t}^{T-1} \xi_s \cdot  \Delta S_s = \pi^\uparrow_t >0, \\
&\text{while on the set}\; A_t\cap \{\pi^\uparrow_t >\pi^\downarrow_t\}\;  \text{we get}\;
&&\delta_{t,T}+ \sum_{s=t}^{T-1} \xi_s \cdot  \Delta S_s >\pi^\downarrow_t \ge 0,
\end{align*}
so that we conclude
\begin{align} \label{temp30}
     	\Ind_{A_t} \big(\delta_{t,T}+ \sum_{s=t}^{T-1} \xi_s \cdot  \Delta S_s \big) \in L^0_+\setminus \{0\}.
\end{align}

At $t$ we take the uniform portfolio strategy $\psi_t^{n} = \Ind_{A_t}\,(n^{-1},...,n^{-1},0,...)$, restricted to the set $A_t$ and $\psi_s^{n}=0$ for all $s\neq t<T$. This is a bounded admissible strategy and since
$$
\sum_{i\ge 1} \frac{1}{i^2} \Var(X_{t,T}^i\,|\,\ccH_{t,T}) =   \Var(X_{t,T}\,|\,\ccH_{t,T}) \sum_{i \ge 1} \frac{1}{i^2}< \infty,
$$
we are entitled to apply the conditional strong law of large numbers given in  Theorem 3.5 in \cite{Majerek2005}. Hence with Assumption \ref{AssX1} and the fact that $\gamma_t=\sum_{i \ge 1} \psi_t^{n,i}=\Ind_{A_t}\in \ccF_t$, we get
    \begin{align}
    	\sum_{i \ge 1} \psi^{n,i}\;X_{t,T}^i \to \Ind_{A_t} E_P[X_{t,T}\,|\,\ccH_{t,T}],
    \end{align}
$P$-almost surely as $n \to \infty$.
With the sub-hedging  strategy $\xi$ from above and the allocation $\psi_t^{n}$ we find
\begin{align*}
  \lim_{n \to \infty} V^I_T(\psi^n)+ V_T^F(\xi) &= \Ind_{A_t}\;\Big(p_t - E[X_{t,T}|\ccH_{t,T}]+ \sum_{s=t}^{T-1} \xi_s \cdot  \Delta S_s
\Big)
\\ &\ge\Ind_{A_t}\;\Big(\delta_{t,T}+ \sum_{s=t}^{T-1} \xi_s \cdot  \Delta S_s\big) \in L^0_+\setminus \{0\}
\end{align*}
by Equation \eqref{temp30}.
The existence of an insurance-finance arbitrage with bounded portfolio strategy is proved.
\end{proof}

The fact that there exists in general a gap between the insurance-finance-consistency of all $\QcirP$-measures and the possibility of an insurance-finance arbitrage, was already shown in Example \ref{exa-discr}.
The reason for this is that the martingale property implied by absence of arbitrage on the financial assets allows for arbitrary equivalent changes of measures on non-traded assets, like the insurance quantities considered here and therefore provides only weak guidance towards efficient pricing. The QP-rule however links the pricing of non-traded wealth to the statistical measure $P$ and therefore produces more reasonable prices.

Now, we make the additional assumption --- interesting in itself --- that the insurer is not able
to exploit arbitrage on the financial market by the information contained in $\GG$, i. e. 
\begin{align}\label{de-NFAZ2}
  \ccM_{e,b}(S,\GG)\neq \emptyset.
    \end{align}
Note that if $Q\in \ccM_{e,b}(S,\GG)$ then the QP-measure $\QcirP$ inherits the Assumption \ref{AssX1*} from the Assumption \ref{AssX1} for the measure $P$.
  \begin{corollary}
  \label{col:NIFA}
Assume that \eqref{de-NFAZ2} holds
and consider  the insurance-finance market $(X,p,S)$ with Assumption \ref{AssX1}. Then, 	 \begin{enumerate}
  \item\label{col:NIFA1} If there exists $Q \in \ccM_{e,b}(S,\GG)$ so that 
\begin{align}  \label{p-cond-A1}
    p_t \le  E_Q\big[ E_P[X_{t,T}|\ccH_{t,T}]|\ccG_t\big] \qquad \text{a.s. for all }\; t<T,
\end{align}
then there is no insurance-finance arbitrage with bounded portfolio strategies ($NIFA^\infty$).
  \item\label{col:NIFA2} If there exists $t<T$, so that $p_t- E_P[X_{t,T}|\ccH_{t,T}]$ is bounded from above and the event that
\begin{align}  
  &p_t > \underset{\scriptscriptstyle{Q \in \ccM_{e,b}(S,\GG)}}{\essinf} E_Q[  E_P[X_{t,T}\,|\,\ccH_{t,T}]  \,|\,\ccG_t]\hspace{15mm} \emph{and}\label{p-cond-A21}\\
  & p_t \ge \underset{\scriptscriptstyle{Q \in \ccM_{e,b}(S,\GG)}}{\esssup} E_Q[  E_P[X_{t,T}\,|\,\ccH_{t,T}] \,|\,\ccG_t]\label{p-cond-A22}
\end{align}
has positive probability under $P$, 
then there exists an insurance-finance arbitrage (IFA) with bounded portfolio strategies.
\end{enumerate}
  \end{corollary}
  
\begin{proof}
\eqref{col:NIFA1} is an immediate consequence of the part \eqref{FTIFA3} $\Rightarrow$\eqref{FTIFA4} of Theorem \ref{thm-FTIFA}, combined with the remark after \eqref{de-NFAZ2}.

The proof of \eqref{col:NIFA2} is similar to the one of Theorem \ref{thm-NIFA}:  $\pi^\downarrow_t:=\underset{\scriptscriptstyle{Q \in \ccM_{e,b}(S,\GG)}}{\essinf}\big(p_t- E_Q[  E_P[X_{t,T}\,|\,\ccH_{t,T}]\big) $ is the sub-hedging price of $p_t- E_Q[  E_P[X_{t,T}\,|\,\ccH_{t,T}]$ which is bounded from above. Restricted to the event $A_t$ defined in \eqref{p-cond-A22}, we get a sub-hedging strategy $\xi =\Ind_{A_t}\;(\xi_t,...,\xi_{T-1})$
so that
\begin{align} \label{temp32}
     \Ind_{A_t}\,\big(p_t- E_Q[  E_P[X_{t,T}\,|\,\ccH_{t,T}]\big)+	  \sum_{s=t}^{T-1} \xi_s \cdot \Delta S_s  &\ge \Ind_{A_t}\pi^\downarrow_t.
\end{align}
Again we have with $\pi^\uparrow_t:=\underset{\scriptscriptstyle{Q \in \ccM_{e,b}(S,\GG)}}{\esssup}\big(p_t- E_Q[  E_P[X_{t,T}\,|\,\ccH_{t,T}] \,|\,\ccG_t]\big)$
\begin{align*}
&\text{on the set}\; A_t\cap \{\pi^\uparrow_t =\pi^\downarrow_t\}\; \text{we have }\;
&& p_t- E_Q[  E_P[X_{t,T}\,|\,\ccH_{t,T}]+ \sum_{s=t}^{T-1} \xi_s \cdot  \Delta S_s = \pi^\uparrow_t >0, \;\text{by}\;\eqref{p-cond-A22}\\
&\text{while on}\; A_t\cap \{\pi^\uparrow_t >\pi^\downarrow_t\}\;  \text{we get}\;
&&p_t- E_Q[  E_P[X_{t,T}\,|\,\ccH_{t,T}]+ \sum_{s=t}^{T-1} \xi_s \cdot  \Delta S_s >\pi^\downarrow_t \ge 0\;\text{by}\;\eqref{p-cond-A21},
\end{align*}
which implies that
\begin{align} \label{temp33}
     	\Ind_{A_t} \big(p_t- E_Q[  E_P[X_{t,T}\,|\,\ccH_{t,T}]\big)+ \sum_{s=t}^{T-1} \xi_s \cdot  \Delta S_s  \in L^0_+\setminus \{0\}.
\end{align}
Now, the construction of a bounded portfolio strategy leading to an insurance-finance arbitrage is identical to the one given in the proof of Theorem \ref{thm-NIFA}.
\end{proof}

\section{Insurance-finance consistency of annuity contracts}\label{sec-annui}
In this section we consider annuity contracts  in order to illustrate the application of our results. We retain the situation of Section \ref{sec-QP-NIFA} where we had the nested sequence of filtrations
$$ \ccF_t\;\subset\;\ccG_t\;\subset\;\ccH_t \qquad\text{for all}\; t\in \TT.$$
and a finance consistent measure on the filtration $\GG$;
\begin{align*}
  \ccM_{e,b}(S,\GG)\neq \emptyset.
    \end{align*}
We begin by inquiring the situation of a progressive enlargement of the financial filtration $\GG$.  We highlight explicitly that $\ccG_0$ is not assumed to be trivial.

\subsection{Progressive enlargement}\label{ssec-progenlarg}

We want to introduce an additional structure on $\GG$ by utilizing the theory of progressive enlargements. For a detailed study of this theory and many references to related literature see \cite{AksamitJeanblanc} and \cite{blanchet-jeanblanc2020}.

In this regard, let
$$G=\big(G_{t,i}\big)_{1\le i\le n_t, t\in \TT}$$
be a nested sequence of finite partitions of $\Omega$, meaning that we have
\begin{flalign} \label{pa-disj}
(i)\qquad&&       &G_{t,i}\cap G_{t,j}=\emptyset \quad \text{for all }\quad t\in \TT,\, i \neq j\le n_t,\quad \text{and}&\notag\\
(ii)\qquad&&          &G_{t,i}=\bigcup \big\{G_{t+1,j}\,\big|\,G_{t+1,j}\subset G_{t,i}\big\} \quad \text{for all } t<T,\, i \le n_t.
    \end{flalign}
Without loss of generality we may assume that $\bigcup_{i\le n_0} G_{0,i}=\Omega$. The filtration
\begin{align} \label{de-G-enl}
      \widetilde \GG=\big( \widetilde\ccG_t\big)_{t\in \TT}\quad \text{with}\quad   \widetilde\ccG_t:=\ccG_t\vee \big\{G_{t,i}\,|\, i \le n_t\big\}\subset \ccH_t  
    \end{align}
is called the \emph{progressive enlargement} of $\GG$ under the partition sequence $G$ and we provide an example in Section \ref{ssec-annui} below.

Note that in this setting, for any $\widetilde\ccG_t$-measurable random variable $Y_t$ we find $\ccG_t$-measurable random variables $Y_{t,i}$, $1\le i \le n_t$, such that $Y_t=\sum_{i\le n_t}\Ind_{G_{t,i}}\,Y_{t,i}$. Therefore, a (discounted) $\widetilde\GG$-adapted financial flow $X=(X_t)_{1\le t\le T}$ (here we write simply $X_t$ instead of $X_{t,T}$ as in previous Sections) can be written in the form
\begin{align} \label{G-flow}
       X_t=\sum_{i=1}^{n_t}\Ind_{G_{t,i}}\,X_{t,i}
    \end{align}
with $\ccG_t$-measurable $X_{t,i}$, $i\le n_t$. Our goal is to value this flow $X$ in an insurance-finance-arbitrage-free way by applying the QP-rule. To this end, we define the densities
\begin{align} \label{de-Lti}
       L_{t,i}:=\QcirP(G_{t,i}|\,\ccG_t),\quad t\in \TT,\, i \le n_t.
\end{align}
The conditional expectation of the flow $X$ under the $\QcirP$-measure can now be calculated as follows:
\begin{proposition}\label{prop-EX}
Let $\widetilde\GG$ be the progressive enlargement of $\GG$ under $G$.
Then, for any $\widetilde\GG$-adapted process $X$, bounded from below, we have
\begin{align}\label{fo-EX}
	E_{\subQcirP}\big[\sum_{1\le t \le T}X_t  \big|\widetilde\ccG_0\big] &= \sum_{i \le n_0}\Ind_{G_{0,i}\cap \{L_{0,i}>0\}}\;L_{0,i}^{-1}
\sum_{1\le t \le T}\sum_{1\le j \le n_t} E_Q\big[ X_{t,j}\,L_{t,i}\big|\ccG_0\big].
\end{align}
\end{proposition}

\begin{proof}
	It suffices to prove \eqref{fo-EX} on each set $G_{0,i}\cap \{L_{0,i}>0\}$, $i\le n_0$. There we have
\begin{align*} 		
	&\Ind_{G_{0,i}\cap \{L_{0,i}>0\}}\;L_{0,i}\,
	E_{\subQcirP}\big[\sum_{1\le t \le T}X_t  \big|G_{0,i}\cap\ccG_0\big]\\
 &\hspace{2cm}= \Ind_{G_{0,i}\cap \{L_{0,i}>0\}}\,E_{\subQcirP}\big[\sum_{1\le t \le T}X_t  \big|\ccG_0\big]\\
  &\hspace{2cm}= \Ind_{G_{0,i}\cap \{L_{0,i}>0\}}\sum_{1\le t \le T}\sum_{1\le j \le n_t}\,E_{\subQcirP}\big[X_{t,j}\Ind_{G_{t,j}}  \big|\ccG_0\big].
  \end{align*}
  Next,
  \begin{align*}
  E_{\subQcirP}\big[X_{t,j}\Ind_{G_{t,j}}  \big|\ccG_0\big]
  &= E_{\subQcirP}\Big[E_{\subQcirP}\big[X_{t,j}\,\Ind_{G_{t,j}} \big|\ccG_t\big]\Big|\ccG_0\Big]
   = E_{\subQcirP}\Big[X_{t,j}\,\QcirP(G_{t,j}  \big|\ccG_t)\Big|\ccG_0\Big]\\
  &= E_{\subQcirP}\big[X_{t,j}\,L_{t,j}\big|\ccG_0\big]
\end{align*}
which shows \eqref{fo-EX}.
\end{proof}

\subsection{An annuity contract}\label{ssec-annui}
We consider a \emph{standard contract of annuity type} issued at date $0$. Besides the information contained in $\ccG_0$, the benefits of the contract depend on the events of death, surrender, or survival up to date $T$.

We recall that $\ccG_0$ is not trivial. 
The benefits are the death benefits (DB), the surrender benefits (SB), or the accumulated benefits (AB), see \cite{ballotta2019variable} for further references and details. 
Let $\sigma$ and $\tau$ denote the random times of surrender and death, respectively and assume that $\sigma >0$, $\tau >0$, to avoid trivialities.

The (already discounted) \emph{death benefit} at time $t$, say $X^1_{t} \in L^0(\Omega,\ccG_t,P)$, is paid if $\tau = t$ and $\sigma > t$. Hence, it is of the form
\begin{align*}
	\DB_{t}= \ind{t=\tau<\sigma\wedge T }\, X_{t}^1.
\end{align*}
The (discounted) \emph{surrender benefit} at time $t<T$, say $X^2_{t} \in L^0(\Omega,\ccG_t,P)$, is paid if $\sigma=t$ and $\tau \ge \sigma$; and the (discounted) \emph{accumulation benefit}, say $X^3_{T} \in L^0(\Omega,\ccG_T,P)$ is paid at $T$ if $\sigma,\tau \ge T$ leading to 
\begin{align*} 
 \SB_{t}&= \ind{t=\sigma\le\tau, t<T}\, X_{t}^2\\
 \AB_{T}&= \ind{T\le\sigma\wedge\tau}\, X_{T}^3    
\end{align*}
The flow of the total benefit $B=(B_t)_{1 \le t \le T}$ accumulates the individual benefits and is given by
\begin{align} \label{t-bene}
  B_{t}=\DB_{t}+\SB_{t}+\AB_{t}.
\end{align}
According to the relevant events of survival and death we build up the partition $G$. Intuitively we decide at time $t$ if death occurred, surrender, or none of these. This leads to the following: $n_0=1$ and $G_{0,1} = \Omega$. Moreover, for $1 \le t \le T$, we set $G_{t,j}= G_{t-1,j},$ for $1\le j\le 2(t-1)$, and 
\begin{align}
     G_{t,2t+1}&= G_{t-1,2t+1} \cap \{t=\tau<\sigma\wedge T\},&\notag\\
     G_{t,2t+2}&= G_{t-1,2t+1}\cap \{t=\sigma\le \tau\},&\notag\\
     G_{t,2t+3}&= G_{t-1,2t+1}\cap \{t<\sigma\wedge \tau\}.
\end{align}
This partition allows us to rewrite the benefits as follows:
\begin{align} \label{beneB}
\DB_{t}&=\Ind_{G_{t,2t+1}}\, X_{t}^1,&\notag\\
\SB_{t}&=\Ind_{G_{t,2t+2}}\, X_{t}^2,&\notag\\
\AB_{T}&=\Ind_{G_{T,2T+1}\cup G_{T,2T+2} \cup G_{T,2T+3}}\, X_{T}^3,
    \end{align}
  and we obtain the following valuation rule for a measure $Q\in \ccM_{e,b}(S,\GG)$.
\begin{proposition}
	\label{cor-annui}
A premium $p_0$ for the flow of benefits described in \eqref{t-bene} satisfies $NIFA ^\infty$ if  
	\begin{align} \label{ineq-annui}
	p_{0} \le \sum_{t=1}^{T-1}	E_Q\Big[ X_{t}^1\,L_{t,2t+1} +X_{t}^2\,L_{t,2t+2} +X_{T}^3\,L_{T-1,2T-1}\Big|\ccG_0\Big].
    \end{align}
\end{proposition}
\begin{proof}
	From Proposition \ref{prop-EX} we obtain
\begin{align} \label{QP-bene}
E_{\subQcirP}\big[\DB_{t} \big|\ccG_0\big] & = \Ind_{G_{0}\cap \{L_{0}>0\}}\;L_{0}^{-1}\;
E_Q\big[ X_{t}^1\,L_{t,2t+1}\big|\ccG_0\big],&\notag\\
E_{\subQcirP}\big[\SB_{t} \big|\ccG_0\big] &= \Ind_{G_{0}\cap \{L_{0}>0\}}\;L_{0}^{-1}\;
E_Q\big[ X_{t}^2\,L_{t,2t+2}\big|\ccG_0\big],&\notag\\
E_{\subQcirP}\big[\AB_{T} \big|\ccG_0\big] &=\Ind_{G_{0}\cap \{L_{0}>0\}}\;L_{0}^{-1}\;
E_Q\big[ X_{T}^3\,L_{T-1,2T-1}\big|\ccG_0\big],&
    \end{align}
where again $L_{0}=\QcirP(G_{0}|\,\ccG_0)$ and $L_{t,j}=\QcirP(G_{t,j}|\,\ccG_t)$ for  $1\le j \le 2t+1$, and $1\le t\le T$.
From the condition \eqref{p-cond-A1} of the Corollary \ref{col:NIFA} we get the concluding result.
\end{proof}

\begin{remark}
	It is possible to extend the setting to the case where  $\widetilde\ccG_0$ is an enlargement of $\ccG_0$ by the finite partition
$G_0=\{G_{0,1},...,G_{0,n_0}\}$ of $\Omega$ according to \eqref{de-G-enl}. To avoid lengthy formulas, we sticked to the simpler case above.
\end{remark}

\section{Conclusion} \label{sec-conclu}
In analogy to the Fundamental Theorem of Asset Pricing (FTAP), the paper defines and investigates arbitrage in the more challenging situation of an insurance company which besides its large portfolio of contracts has the possibility to hedge its risks in a financial market. The corresponding Fundamental Theorem provides so called insurance-finance-consistent measures, but unfortunately they lack the connection to the statistical data of the company. The QP-rule is the state of the art for this connection; consequently its insurance-finance consistency has to be analysed.

In the context of a linear relation between premiums and benefits our work can be seen as the first step towards a general study of the consistency of insurance flows. Future work will be needed to include non-linear entities, like risk measures and solvency capital requirements, and should take the possible regulated transfer of insurance risk from a ruined company to another into account.

\begin{appendix}
\section{The Key Proposition}
The following considerations based on ideas in Chapter 6 of \cite{DelbaenSchachermayer2006}. See also \cite{schachermayer1992} where the projection method below appeared for the first time. Similar results are given in \cite{KabanovStricker2006}. For the convenience of the reader we provide a complete proof of the Key Proposition \ref{aux3}. For some readers it may be interesting in its own right.

For $t=0,...,T-1$ we define backward recursively the following $\ccG_t$-stable subspace of $L^0(\ccG_t)$: First we have the kernel of the linear mapping $\xi_t \mapsto xi_t\cdot \Delta S_t$:
\begin{align*}
\widetilde{\kN}_t:=\big\{\xi_t\in L^0(\ccG_t)\;|\; \xi_t\cdot \Delta S_t=0\big\}.
\end{align*}
By Lemma 6.2.1. in \cite{DelbaenSchachermayer2006} we get a $\ccG_t$-measurable projection
$\widetilde{\rho}_t$ so that
\begin{align*}
\xi_t\in \widetilde{\kN}_t \Longleftrightarrow \widetilde{\rho}_t(\xi_t)=\xi_t.
\end{align*}
The orthogonal complement is
\begin{align*}
\widehat{\kN}_t:=\big\{\xi_t\in L^0(\ccG_t)\;|\; (Id-\widetilde{\rho}_t)(\xi_t)=\xi_t\big\}.
\end{align*}
For $t=T-1$ we set $\kN_{T-1}=\widehat{\kN}_{T-1}$. For $t<T-1$,  assume that  the sets $\kM_s$ and $\kN_s$ are already defined for $t<s\le T-1$. We consider the linear mapping
\begin{align*}
&\kJ_t: \bigtimes_{t<s\le T-1}\big( \kM_s\times \kN_s\big) \longrightarrow L^0(\ccH_{T-1,T}),\\
&\kJ_t\big((\gamma_s,\xi_s)_{t<s\le T-1}\big)=\sum_{t<s\le T-1}\gamma_s \cdot Y_{s,T}+\xi_s\cdot\Delta S_s.
\end{align*}
and we have the extended kernel
\begin{align*}
\widetilde{\widetilde{\kN}}_t:=(\kJ_t)^{-1}
\Big(\kJ_t\big(\bigtimes_{s=t+1}^{T-1}( \kM_s\times \kN_s)\big)
\cap \big\{\xi_t\cdot \Delta S_t\big|\;\xi_t\in L^0(\ccG_t)\big\}\Big).
\end{align*}
Again let
$\widetilde{\widetilde{\rho}}_t$ be the $\ccG_t$-measurable projection on $\widetilde{\widetilde{\kN}}_t$:
\begin{align*}
\xi_t\in \widetilde{\widetilde{\kN}}_t \Longleftrightarrow \widetilde{\widetilde{\rho}}_t(\xi_t)=\widetilde{\widetilde{\rho}}_t
\big(Id-\widetilde{\rho}_t\big)(\xi_t)=0.
\end{align*}
Finally, the orthogonal complement of the extended kernel is
\begin{align*}
\kN_t:=\big\{\xi_t\in L^0(\ccG_t)\;|\; (Id-\widetilde{\widetilde{\rho}}_t)(Id-\widetilde{\rho}_t)(\xi_t)
=\xi_t\big\}.
\end{align*}
We now turn to the analogous definitions of $\kM_t$, starting with the kernel
\begin{align*}
\widetilde{\kM}_t:=\big\{\gamma_t\in L^0_+(\ccG_t)\;|\; \gamma_t\cdot Y_{t,T}=0\big\},
\end{align*}
and let $\widetilde{\pi}_t$ be the projection on $\widetilde{\kM}_t$ with
\begin{align*}
\gamma_t\in \widetilde{\kM}_t \Longleftrightarrow \widetilde{\pi}_t(\gamma_t)=\gamma_t
\end{align*}
and the orthogonal complement
\begin{align*}
\widehat{\kM}_t:=\big\{\gamma_t\in L^0(\ccG_t)\;|\; (Id-\widetilde{\pi}_t)(\gamma_t)=\gamma_t\big\}.
\end{align*}
Here, we consider the linear mapping
\begin{align*}
&\kI_t:  \kN_t \times \bigtimes_{t<s\le T-1}\big( \kM_s\times \kN_s\big) \longrightarrow L^0(\ccH_{T-1,T}),\\
&\kI_t\big(\xi_t, (\gamma_s,\xi_s)_{t<s\le T-1}\big):=\xi_t\cdot\Delta S_t+\sum_{t<s\le T-1}\gamma_s \cdot Y_{s,T}+\xi_s\cdot\Delta S_s.
\end{align*}
Note that for $t=T-1$ the map $\kI_{T-1}$ id defined only on $\kN_{T-1}$ which is defined above. We consider the extended kernel
\begin{align*}
\widetilde{\widetilde{\kM}}_t:=(\kI_t)^{-1}
\Big(\kI_t\big(\kN_t\times \bigtimes_{s=t+1}^{T-1}( \kM_s\times \kN_s)\big)
\cap \big\{\gamma_t\cdot Y_{t,T}\big|\;\gamma_t\in L^0
_+(\ccG_t)\big\}\Big).
\end{align*}
entailing the projection $\widetilde{\widetilde{\pi}}_t$ on $\widetilde{\widetilde{\kM}}_t$:
\begin{align*}
\gamma_t\in \widetilde{\widetilde{\kM}}_t \Longleftrightarrow \widetilde{\widetilde{\pi}}_t\big(Id-\widetilde{\pi}_t\big)
(\gamma_t)=\gamma_t
\end{align*}
Finally, its orthogonal complement is
\begin{align*}
\kM_t:=\big\{\gamma_t\in L^0(\ccG_t)\;|\; (Id-\widetilde{\widetilde{\pi}}_t)(Id-\widetilde{\pi}_t)(\gamma_t)
=\gamma_t\big\}.
\end{align*}
Note that $\ind{Y_{t,T}=0}\;\kN_t=\{0\}$ and $\ind{\Delta S_t=0}\;\kM_t=\{0\}$.\\
We say that  $\big((\gamma_t,\xi_t)\big)_{0\le t \le T-1}$ is in \emph{canonical form} if $\gamma_t\in \kM_t$ and $\xi_t\in \kN_t$ for all $0\le t \le T-1$.
As a first result we have the following Proposition:
\begin{proposition}\label{aux1}
  The mapping
\begin{align*}
&\kJ: \bigtimes_{0\le t\le T-1}\big( \kM_t\times \kN_t\big) \longrightarrow L^0(\ccH_{T-1,T}),\\
&\kJ\big((\gamma_t,\xi_t)_{0 \le t\le T-1}\big)\;:=\;\sum_{0 \le t\le T-1}\gamma_t \cdot Y_{t,T}+\xi_t\cdot\Delta S_t.
\end{align*}
is injective.
\end{proposition}
\begin{proof}
 Set $f_t= \sum_{t \le s\le T-1}\gamma_s \cdot Y_{s,T}+\xi_s\cdot\Delta S_s$ and assume that $t_0$ is the maximal time with $f_0=f_{t_0}$. Then for $s< t_0$ we have $f_0\in \kJ_s\big(\bigtimes_{s<\tau\le T-1}\big( \kM_\tau\times \kN_\tau\big) \big) \cap \kI_s\big(\{0\}\times \bigtimes_{s<\tau\le T-1}\big( \kM_\tau\times \kN_\tau\big)\big)$ such that $\gamma_s\in \widetilde{\widetilde{\kM}}_s \cap \kM_s =\{0\}$ and $\xi_s\in \widetilde{\widetilde{\kN}}_s \cap \kN_s =\{0\}$. This shows $(\gamma_s,\xi_s)=(0,0)$ for $s<t_0$.
  Next assume that $f_{t_0} \in \kI_t\big(\kN_{t_0} \times \bigtimes_{{t_0}<s\le T-1}( \kM_s\times \kN_s)\big)$ which implies $\gamma_{t_0}\in \widetilde{\widetilde{\kM}}_{t_0}\cap \kM_{t_0}=\{0\}$, hence $\gamma_{t_0}=0$ again. Otherwise $\gamma_{t_0}\cdot Y_{t_0,T}\neq 0$ and $\gamma_{t_0}\in \kM_{t_0}$ is unique. Now regard $\widetilde{f}_{t_0}=\xi_{t_0}\cdot\Delta S_{t_0}+ f_{t_0+1}$. If
  $\widetilde{f}_{t_0}\in \kJ_t\big(\bigtimes_{t<s\le T-1}( \kM_s\times \kN_s)$ then $\xi_{t_0}\in \widetilde{\widetilde{\kN}}_{t_0}\cap \kN_{t_0}=\{0\}$, i. e. $\xi_{t_0}=0$.  Otherwise $\xi_{t_0}\cdot \Delta S_{t_0}\neq 0$ and $\xi_{t_0}\in \kN_{t_0}$ is unique. The uniqueness of $(\gamma_s,\xi_s)$ for $s>t_0$ follows by the induction hypothesis.
  \end{proof}
Another preliminary result is the following Proposition:
\begin{proposition}\label{aux2}
  Assume condition \eqref{FTIFA2} of Theorem \ref{thm-FTIFA} and  $(\gamma^n,\xi^n)$ is a sequence of canonical form, i. e. $(\gamma^n_t,\xi^n_t)\in \kN_t \times \kM_t$. Then we have:
    \begin{align*}
(\gamma^n,\xi^n) \text{is bounded a.s.}\quad\Longleftrightarrow\quad
\Big(\sum_{t=0}^{T-1}\big(\gamma^n_t\cdot Y_{t,T} + \xi^n_t \cdot \Delta S_t\big)\Big)_- \quad\text{is bounded a.s.}
\end{align*}
\end{proposition}
\begin{proof}
The $\Longrightarrow$-part is trivial. The inverse direction is proved by induction on $T$. For $T=0$ there is nothing to show. For $T\ge 1$ we regard a sequence
\begin{align}\label{seq-ga-xi}
(\gamma^n, \xi^n)=\big((\gamma^n_0, \xi^n_0),...(\gamma^n_{T-1}, \xi^n_{T-1})\big)
\end{align}
and set for $0\le t\le T-1$
\begin{align}\label{seq-V-t-n}
V^n_t\;:= \sum_{\tau=t}^{T-1}\big(\gamma^n_\tau\cdot Y_{\tau,T} + \xi^n_\tau \cdot \Delta S_\tau\big).
\end{align}
First assume that the set
\begin{align*}
A:=\big\{ \liminf (\gamma^n_0+\norm{\xi^n_0}) = \infty \big\}\in\ccG_0
\end{align*}
has positive measure $P(A)>0$.
We divide the set $A$ into two subsets:
\begin{align*}
& A_1 :=A\cap \big\{ \liminf \norm{\xi^n_0}\big/\gamma^n_0 =0,\; \gamma^n_0\ge 1 \big\},\\
& A_2 :=A\cap \big\{ \liminf \gamma^n_0\big/\norm{\xi^n_0}<\infty,\; \norm{\xi^n_0}\ge 1 \big\}.
\end{align*}
Assume first that $P(A_1)>0$. Note that $Y_{0,T} \neq 0$ on $A_1$ since the $\gamma^n_0>0$  on $A_1$ are in canonical form. By the measurable selection principle (see \cite{DelbaenSchachermayer2006} Proposition 6.3.3.) we find a $\ccG_0$-measurable subsequence $\tau^n$ so that $\gamma^{\tau^n}_0\rightarrow \infty$ and $\norm{\xi^{\tau^n}_0}\big/\gamma^{\tau^n}_0\rightarrow 0$ on the set $A_1$. Then
\begin{align*}
&\Ind_{A_1}\;Y_{0,T}=\Ind_{A_1}\;\lim_n \big( \gamma^{\tau^n}_0\cdot Y_{0,T} + \xi^{\tau^n}_0 \cdot \Delta S_0\big)\Big/\gamma^{\tau^n}_0 \hspace{60mm}\quad\text{and}\\
&\Ind_{A_1}\;\limsup_n\; \frac{\big(V^{\tau^n}_1\big)_-}{\gamma^{\tau^n}_0}
\le
 \Ind_{A_1}\; \limsup_n \frac{\big(V^{\tau^n}_0\big)_-}{\gamma^{\tau^n}_0}+\limsup_n \Big(\big( \gamma^{\tau^n}_0\cdot Y_{0,T} + \xi^{\tau^n}_0 \cdot \Delta S_0\big)_-\big/\gamma^{\tau^n}_0\Big)\\
 &\hspace{33mm} \le \abs{Y_{0,T}}
\end{align*}
since $\Ind_{A_1}\;\limsup_n \big(\big(V^{\tau^n}_0\big)_-\big/\gamma^{\tau^n}_0\big)=0$ by  assumption. Hence, $\big(V^{\tau^n}_1\big)_-\big/\gamma^{\tau^n}_0$ is bounded on $A_1$  and so is $\Ind_{A_1}\;\big((\gamma^{\tau^n}_1, \xi^{\tau^n}_1),...(\gamma^{\tau^n}_{T-1}, \xi^{\tau^n}_{T-1})\big)\big/\gamma^{\tau^n}_0$ by induction hypothesis. By a number of applications of the measurable selection principle we get series of iterated subsequences $\sigma^n_1$ up to $\sigma^n_{T-1}$ so that for $1\le t\le T-1$ we have
\begin{align*}
\Ind_{A_1}\;\lim_n (\gamma_t^{\sigma_t^n},\xi_t^{\sigma_t^n})/\gamma_0^{\sigma_t^n}=
\Ind_{A_1}\;\lim_n (\gamma_t^{\sigma_{T-1}^n},\xi_t^{\sigma_{T-1}^n})
/\gamma_0^{\sigma_{T-1}^n}
=(\widetilde{\gamma}_t,\widetilde{\xi}_t)
\end{align*}
for some $(\widetilde{\gamma}_t,\widetilde{\xi}_t)\in L^0_+(\ccG_t)\times L^0(\ccG_t)$.  Therefore,
\begin{align*}
\Ind_{A_1}\;\lim_n V^{\sigma_{T-1}^n}_0/\gamma_0^{\sigma_{T-1}^n}= \Ind_{A_1}\;\Big(Y_{0,T}+\sum_{\tau=1}^{T-1}\big(\widetilde{\gamma}_\tau\cdot Y_{\tau,T} + \widetilde{\xi}_\tau \cdot \Delta S_\tau\big)\Big)=:\Ind_{A_1}\;Y_{0,T}+\widetilde{V}_1.
\end{align*}
While we till have $\Ind_{A_1}\;\lim_n \Big(Y_{0,T}+V^{\sigma_{T-1}^n}_1/\gamma_0^{\sigma_{T-1}^n}\Big)_-\;=0$ it follows that $\Ind_{A_1}\;Y_{0,T}+\widetilde{V}_1\ge 0$ and  condition \eqref{FTIFA2} of Theorem \ref{thm-FTIFA} implies that $\Ind_{A_1}\;Y_{0,T}+\widetilde{V}_1= 0$. This contradicts Proposition \ref{aux1} according to which the coefficient of $Y_{0,T}$ must be $0$.\medskip\\
Next let $P(A_2)>0$. Here, the selection principle yields a $\ccG_0$-measurable subsequence $\tau^n$ so that $\norm{\xi^{\tau^n}_0}\rightarrow \infty$, $\lim_n \xi^{\tau^n}_0/\norm{\xi^{\tau^n}_0}=\xi_0$ with $\norm{\xi_0}=1$, and $\lim_n \gamma^{\tau^n}_0\big/\norm{\xi^{\tau^n}_0}=:\gamma_0< \infty$ on the set $A_2$. We define
\begin{align*}
&\widetilde{\xi}_0:= \Ind_{A_2}\;\xi_0 \in \kN_0\setminus\{0\} \hspace{90mm}\quad\text{and}\\
&\widetilde{\gamma}_0:\Ind_{A_2}\; \gamma_0 \in \kN_0.
\end{align*}
With similar arguments as above, we have
 \begin{align*}
&\widetilde{\gamma}_0\cdot Y_{0,T}+\widetilde{\xi}_0\cdot \Delta S_0 =\Ind_{A_2}\;\lim_n \big( \gamma^{\tau^n}_0\cdot Y_{0,T} + \xi^{\tau^n}_0 \cdot \Delta S_0\big)\Big/\norm{\xi^{\tau^n}_0} \hspace{35mm}\quad\text{and}\\
&\Ind_{A_2}\;\limsup_n\; \frac{\big(V^{\tau^n}_1\big)_-}{\norm{\xi^{\tau^n}_0}}\le
 \Ind_{A_2}\; \limsup_n \frac{\big(V^{\tau^n}_0\big)_-}{\norm{\xi^{\tau^n}_0}}+\limsup_n \Big( \gamma^{\tau^n}_0\cdot Y_{0,T} + \xi^{\tau^n}_0 \cdot \Delta S_0\Big)_-\Big/\norm{\xi^{\tau^n}_0}\Big)\\
 &\hspace{33mm} \le \abs{\widetilde{\gamma}_0\cdot Y_{0,T}+\widetilde{\xi}_0\cdot \Delta S_0}
\end{align*}
and by induction hypothesis $\Ind_{A_2}\;\big((\gamma^{\tau^n}_1, \xi^{\tau^n}_1),...(\gamma^{\tau^n}_{T-1}, \xi^{\tau^n}_{T-1})\big)\big/\norm{\xi^{\tau^n}_0}$ is bounded. Again, iterated applications of the selection principle yield elements $(\gamma_t,\xi_t)\in  \Ind_{A_2}\;(\kM_t,\kN_t)$ for $t=1,...,T-1$ with
\begin{align*}
&\Ind_{A_2}\;\lim_n V^{\sigma_{T-1}^n}_0/\norm{\xi^{\sigma_{T-1}^n}}= \widetilde{\gamma}_0\cdot Y_{0,T} +\widetilde{\xi}_0\cdot \Delta S_0+\sum_{\tau=1}^{T-1}\big(\widetilde{\gamma}_\tau\cdot Y_{\tau,T} + \widetilde{\xi}_\tau \cdot \Delta S_\tau\big)\\
&\hspace{33mm}=:\widetilde{V}_0=:\widetilde{\gamma}_0\cdot Y_{0,T} +\widetilde{\xi}_0\cdot \Delta S_0+\widetilde{V}_1.
\end{align*}
Since $\big(\widetilde{V}_0\big)_-=0$, condition \eqref{FTIFA2} of Theorem yields $\widetilde{V}_0=0$ contradicting by Proposition \ref{aux1} the coefficient $\widetilde{\xi}_0$ since $\norm{\widetilde{\xi}_0}=1$.\\
Thus, the sequence $(\gamma^n_0, \xi^n_0)$ is bounded. Once again
\begin{align*}
\big(V^n_1\big)_-\le
\big(V^n_0\big)_- + \big(\gamma^n_0\cdot Y_{0,T} + \xi^n_0 \cdot \Delta S_0\big)_-
\end{align*}
shows that $(V^n_1)_-$ is bounded and induction hypothesis gives us that for $1\le t\le T-1$ the sequences $(\gamma^n_t, \xi^n_t)$ are bounded, too.
\end{proof}
With the last two Propositions we arrive at the main Proposition needed to the last part of the proof of Theorem \ref{thm-FTIFA}.

\begin{proposition}\label{aux3}
  Under condition \eqref{FTIFA2} of Theorem \ref{thm-FTIFA} and with $\ccK$ from \eqref{de-K} the cone
    \begin{align*}
	    	\ccK\;-\; L^0_+(\ccH_{T-1,T})
    \end{align*}
is closed.
\end{proposition}
\begin{proof}
With the notations from \eqref{seq-ga-xi} and \eqref{seq-V-t-n} let $(\gamma^n, \xi^n)$
be a sequence in canonical form so that $V^n_0- H^n \longrightarrow G$ for a sequence $H^n\in L^0_+(\ccH_{T-1,T})$ and $G\in L^0(\ccH_{T-1,T})$. Since $\liminf_n V^n_0\ge G$  we have $\limsup^n\big(V^n\big)_-\ge G_-$ and $\big(V^n\big)_-$ is bounded. So is the sequence $(\gamma^n, \xi^n)$ is bounded by Proposition \ref{aux2}. As in the proof before, iterated applications of the measurable selection principle yield that $(\gamma_t^{\sigma_{T-1}^n},\xi_t^{\sigma_{T-1}^n})\longrightarrow (\gamma_t,\xi_t) \in \kM_t\times \kN_t$ and $\lim_n V^{\sigma_{T-1}^n}_0= \sum_{t=0}^{T-1} \gamma_t\cdot Y_{t,T-1}+\xi_t\cdot \Delta S_t=:V_0$. Then $H^{\sigma_{T-1}^n}\longrightarrow V_0-G \ge 0$ and $V_0-G\in \ccK\;-\; L^0_+(\ccH_{T-1,T})$.
\end{proof}

\section{Evaluation of insurance results}

\begin{proposition} \label{prop-V-mean}
 Under  Assumption \ref{AssX1} and the admissibility condition we have for all bounded portfolio strategies $\psi$, all  $L_{t,T}\in L^\infty(\ccH_{t,T})$ and all $t=0,\dots,T-1$, that
 \begin{align}
 	E_P\Big[ L_{t,T}\,\lim_{n \to \infty} \sum_{i \ge 1} \psi^{n,i}_t \, p_t \Big]&=  E_P[ L_{t,T} \, \gamma_t \, p_t],   & and
 	\label{temp26}\\
 	E_P\Big[ L_{t,T}\, \lim_{n \to \infty} \sum_{i \ge 1} \psi^{n,i}_t X_{t,T}^i \Big]
 		& = E_P\Big[L_{t,T} \, \gamma_t    E_P\big[X_{t,T}\big|\ccH_{t,T}\big]\Big]. &
 	\label{temp27} 	
\end{align}
\end{proposition}

\begin{proof}
Indeed, \eqref{temp26} follows  from the convergence of the insurance volume in \eqref{conv-psi}.

To prove \eqref{temp27}, we first show that $(\sum_{i\ge 1}\psi_t^{n,i}X_{t,T}^{i})_{n \in \NN}$ is uniformly integrable.
It holds
\begin{align}\label{eq:uniforml2bound1}
	E_P\bigg[  \Big(\sum_{i \ge 1} \psi_t^{n,i}X_{t,T}^i\Big)^2 \bigg]
	&=  E_P\bigg[E_P\Big[  \Big(\sum_{i \ge 1} \psi_t^{n,i}X_{t,T}^i\Big)^2 \Big\vert \ccH_{t,T}\Big]\bigg] \nonumber \\
	&= E_P\bigg[\sum_{i,j \ge 1}  E_{P}\Big[ \psi_t^{n,i}X_{t,T}^i\psi_t^{n,j}X_{t,T}^j \Big\vert \ccH_{t,T}\Big]\bigg]\nonumber \\
	&=E_P\bigg[\sum_{i,j \ge 1} \psi_t^{n,i}\psi_t^{n,j} E_P\left[ X_{t,T}^iX_{t,T}^j\vert \ccH_{t,T}\right]\bigg].
\end{align}
Assumption \ref{AssX1} yields that
\begin{align*}
\eqref{eq:uniforml2bound1} &\le
E_P\left[\sum_{i\neq j\ge 1}  \psi_t^{n,i}\psi_t^{n,j}E_P\left[ X_{t,T}\vert \ccH_{t,T}\right]^2\right]
+ E_P\left[ \sum_{i \ge 1}(\psi_t^{n,i})^2 E_P\left[ X_{t,T}^2\vert \ccH_{t,T}\right] \right]
<\infty,
\end{align*}
uniformly in $n\in\mathbb N$. Hence, we  obtain uniform integrability of $\sum_{i\ge 1}\psi_t^{n,i}X_{t,T}^{i}$.	
From this, it follows that
\begin{align}
\int  L_{t,T}\, E_P\big[ \lim_{n \to \infty} \sum_{i \ge 1} \psi^{n,i}_t  X_{t,T}^i | \ccH_{t,T} \big] dP
&=
\int  L_{t,T}\, \lim_{n \to \infty}  E_P\big[  \sum_{i \ge 1} \psi^{n,i}_t  X_{t,T}^i | \ccH_{t,T} \big] dP \notag \\
&=
\int  L_{t,T}\, \lim_{n \to \infty}  \sum_{i \ge 1} \psi_t^{n,i} \, E_P\big[ X_{t,T}^i | \ccH_{t,T} \big] dP \notag\\	
&=
\int  L_{t,T}\,  \gamma_t \, E_P\big[ X_{t,T} | \ccH_{t,T} \big] dP,		\label{temp1181}
\end{align}
and the claim is proven.
\end{proof}

Since we will consider $P^* \sim P$, a portfolio strategy $\psi$ is admissible under $P$, iff it is admissible under $P^*$.

\begin{proposition}
\label{prop-V-mean*}  Assume that  Assumption \ref{AssX1} holds (under $P$).
	Consider a measure $P^*$ under which Assumption \ref{AssX1*} is satisfied, let $\psi$ be a  bounded and admissible portfolio strategie $\psi$, and  $t=0,\dots,T-1$. Then, 
 \begin{align}
 	E_{P^*}\Big[  \lim_{n \to \infty} \sum_{i \ge 1} \psi^{n,i}_t X_{t,T}^i \Big]
 		& = E_{P^*}\Big[ \, \gamma_t \,    E_P\big[X_{t,T}\big|\ccH_{t,T}\big]\Big]. 
 	\label{temp27*} 	
\end{align}
\end{proposition}

\begin{proof}
Since $P \sim P^*$ almost sure limit results persist under this change of measure. Since from Assumption \ref{AssX1} we obtain 
	as in the proof of Proposition \ref{prop-V-mean}  that $(\sum_{i\ge 1}\psi_t^{n,i}X_{t,T}^{i})_{n \in \NN}$ is uniformly integrable under $P$, it is uniformly integrable under $P^*$.
	
As in Equation \eqref{temp1181} we obtain
\begin{align}
\int  E_{P^*}\big[ \lim_{n \to \infty} \sum_{i \ge 1} \psi^{n,i}_t  X_{t,T}^i | \ccH_{t,T} \big] dP^*
&=
\int  \lim_{n \to \infty}  \sum_{i \ge 1} \psi_t^{n,i} \, E_{P^*}\big[ X_{t,T}^i | \ccH_{t,T} \big] dP^* \notag\\	
&=
\int  \gamma_t \, E_{P^*}\big[ X_{t,T} | \ccH_{t,T} \big] dP^*.
\end{align}
Finally, we observe that under Assumption \ref{AssX1},  the uniform portfolio strategy 
 \begin{align*}
    \psi^n_t= (1/n,\ldots,1/n,0...),
\end{align*}
 together with the conditional strong law of large numbers in \cite{Majerek2005} gives for this allocation the following result:
 \begin{align*}
     \lim_{n \to \infty} \sum_{i \ge 1} \psi^{n,i}_t   X_{t,T}^i 
        & =   E_P\big[    X_{t,T}\big|\ccH_{t,T}\big] 
\end{align*}
on a measurable set $A$ with $P(A)=1$.
Under $P^*$, we obtain similarly 
 \begin{align*}
     \lim_{n \to \infty} \sum_{i \ge 1} \psi^{n,i}_t   X_{t,T}^i 
        & =   E_{P^*}\big[    X^1_{t,T}\big|\ccH_{t,T}\big] 
\end{align*}
on $A'$ with $P^*(A')=1$. 
 Since $P\sim P^*$ we obtain $E_{P^*}\big[    X^1_{t,T}\big|\ccH_{t,T}\big] = E_{P}\big[    X_{t,T}\big|\ccH_{t,T}\big]$ and the claim is proven.
\end{proof}

\end{appendix}

If the density $L=dP^*/dP$ is $\ccH_{0,T}$-measurable, then \eqref{temp27*} follows also by an Application of Proposition \ref{prop-V-mean}. If, however, $L$ is not $\ccH_{0,T}$-measurable we need Assumption \ref{AssX1*}.

\end{document}